\documentclass[useAMS,usenatbib]{mn2e}
\usepackage{epsfig}
\usepackage{color}
\usepackage{subfigure}
\usepackage{natbib}
\usepackage{aas_macros}

\title[Higher order moments vs MICE]{Higher order moments of lensing convergence - I. Estimate from simulations}

\author[M. Vicinanza et al.]{M. Vicinanza$^{1,2,3}$\thanks{Corresponding author\,: {\tt martina.vicinanza@oa-roma.inaf.it}}, V.F. Cardone$^{1}$, R. Maoli$^{3}$, R. Scaramella$^{1}$, X. Er$^{1}$ \\
$^1$I.N.A.F.\,-\,Osservatorio Astronomico di Roma, via Frascati 33, 00040 - Monte Porzio Catone (Roma), Italy \\
$^2$Dipartimento di Fisica, Universit\`{a} di Roma "Tor Vergata", via della Ricerca Scientifica 1, 00133 - Roma, Italy \\
$^3$Dipartimento di Fisica, Universit\`{a} di Roma "La Sapienza", Piazzale Aldo Moro, 00185 - Roma, Italy \\}

\date{Accepted xxx, Received yyy, in original form zzz}

\begin{document}

\maketitle

\begin{abstract}

Large area lensing surveys are expected to make it possible to use cosmic shear tomography as a tool to severely constrain cosmological parameters. To this end, one typically relies on second order statistics such as the two\,-\,point correlation fucntion and its Fourier counterpart, the power spectrum. Moving a step forward, we wonder whether and to which extent higher order stastistics can improve the lensing Figure of Merit (FoM). In this first paper of a series, we investigate how second, third and fourth order lensing convergence moments can be measured and use as probe of the underlying cosmological model. We use simulated data and investigate the impact on moments estimate of the map reconstruction procedure, the cosmic variance, and the intrinsic ellipticity noise. We demonstrate that, under realistic assumptions, it is indeed possible to use higher order moments as a further lensing probe.

\end{abstract}


\section{Introduction}

Weak gravitational lensing (hereafter, WL) refers to the small distortion of galaxy image due to the combined effect of the matter along the line of sight. The shape change is, however, so small that can only be detected statistically relying on large galaxy samples. In particular, WL induces a correlation between galaxy shapes so that the ellipticity correlation function is altered and bring the signature of the WL effect. Since the distortion depends on both the matter distribution along the line of sight and the underlying cosmic evolution, WL turns out to be an efficient probe of the cosmological parameters motivating the great interest towards this topic as witnessed by many recent reviews (see, e.g., \citealt{B&S2001,Mun2008,Kilby2014}). Moreover, lensing surveys as the CFHTLenS \citep{H12} and RCSLensS \citep{H16} have convincingly demonstrated the techinical feasibility of WL measurements paving the way to larger ongoing surveys such as the Dark Enery Survey \citep{DE
 S2015} and the Kilo Degree Survey \citep{Kids} and future projects both ground based,  as the Large Synoptic Survey Telescope \citep{LSST2009},  and from space, as Euclid \citep{Laur2011} and WFIRST \citep{WFIRST2012}. 

From an observational point of view, all the above quoted surveys aim at mapping cosmic shear from the ellipticity of large galaxy samples. Shear catalogs are then used to determine the two points correlation function splitting sources in redshift bins to perform tomography. This is then contrasted against theoretical expectations thus constraining the cosmological parameters (see, e.g., \citealt{Jee13,Kil13,Jee15,Abb15}). Alternatively, one can rely on a 3D analysis based on spherical Fourier\,-\,Bessel decompisition to constrain both dark energy models \citep{Tom14} and modified gravity theories \citep{Pr16}. It is worth noting that both these techniques mainly probe the growth of structures in the linear and quasi\,-\,linear regime so that the constraints on cosmological parameters are affected by severe degeneracies such as, e.g., the well known $\Omega_M$\,-\,$\sigma_8$ one.

In order to break this degeneracy, a possible way out consists in moving to a different tracer. Indeed, cosmic shear measurements can also be used to reconstruct the convergence field which quantifies the projection along the line of sight of the density contrast $\delta({\bf x}= [\rho({\bf x}) - \bar{\rho}]/\bar{\rho}$. What makes convergence so attractive is the different scales which is sensible to. Indeed, the convergence field mainly probes the nonlinear scales \citep{JSW2000,MJ2000,V2000,MJ01,TJ2004,TW2004,VMB2005} thus offering complementary information to the shear field. Moreover, on these scales, the collapse of structures introduce deviations from the Gaussianity which are strongly related to the underlying cosmological model and theory of gravity. The non Gaussianity of the field can be quantified through higher than second order moments which can then be used to break the $\Omega_M$\,-\,$\sigma_8$ degeneracy \citep{Berny1997,JaSel1997}. Actually, higher order mom
 ents depend on the full set of cosmological parameters so that it is worth investigating whether they can represent a valuable help to narrow down the uncertainties on the dark energy equation of state too. Put in other words, one can wonder whether convergence moments can help to increase the Figure of Merit (FoM) if used alone and/or in combination with the standard second order cosmic shear tomography.

This is the first in a series of papers where we aim at answering the above question considering both observational and theoretical aspects for moments up to the fourth order. In the present paper, we start by addressing the problem of measuring moments and comparing them to theoretical predictions. In particular, we investigate the impact of map reconstruction, cosmic variance and intrinsic ellipiticity noise. To this end, we make extensive use of simulated weak lensing maps from the MICE collaboration \citep{Carr2015,Fos2015a,Crocce2015b,Fos2015b,Hoff15} in order to mimic the output of a realistic future lensing survey

The plan of the paper is as follows. In Sect. 2 we give an overview on the simulated dataset and which kind of analysis we will perform. In Sect. 3 we present the strategy we will adopt to compute the high order moments. In Sect. 4 and Sect. 5 we will improve our measurements mimic a more realistic scenario adding some systematics effects. Finally, in Sect. 6 and Sect. 7 we will discuss our preliminary results and present our conclusions.

\section{The simulated dataset}

N\,-\,body light cone simulations are an ideal tool to build all\,-\,sky lensing maps. In particular, the MICE Grand Challenge simulation \citep{Fos2015a,Crocce2015b}, containing about 70 billion dark matter particles in a $(3 h^{-1} 	\ {\rm Gpc})^3$ volume, has the volume and mass resolution needed to guarantee an accurate modeling of the lensing observables for upcoming wide and deep galaxy surveys. Halo and galaxy catalogs have been built using  a hybrid Halo Occupation Distribution (HOD) and Halo Abundance Matching (HAM) prescription to populate Friends\,-\,of\,-\,Friends (FoF) dark matter haloes \citep{Carr2015} so that local observational constraints, such as the local luminosity function \citep{Blan2003,Blan2005a}, the galaxy clustering as a function of luminosity and colour \citep{Zeha2011} and colour	\,-\,magnitude diagrams \citep{Blan2005b}, are fullfilled. Following the {\it Onion universe} approach \cite{Fos2015b}, a lensing catalog is generated too giving, in eac
 h galaxy position, the values of the convergence $\kappa$ and the two shear components $(\gamma_1, \gamma_2)$ as inferred directly from the dark matter distribution along the line of sight. We use the version 2.0 which extends to less massive and hence lower luminosity galaxies being complet to $i \sim 24$ at all redshifts in the range $(0.1, 1.4)$. We remind the reader that a flat $\Lambda$CDM model is assumed setting the relevant parameters as follows

\begin{displaymath}
(\Omega_M, \Omega_b, h, n_s, \sigma_8) = (0.25, 0.044, 0.70, 0.95, 0.80)
\end{displaymath}
being $(\Omega_M, \Omega_b, h, n_s, \sigma_8) $ the matter and baryon density parameters, the Hubble constant in units of $100 \ {\rm km/s/Mpc}$, the spectral index and the variance of perturbations on the scale $R = 8 h^{-1} \ {\rm Mpc}$, respectively.

We cut 140 $5 \times 5 \ {\rm sq deg}$ non contiguous patches summing up to a total area $\Omega = 3500 \ {\rm sq \ deg}$. Note that the full MICECAT area is actually larger, but we have reduced it in order to have well separated fields all with the same approximately square shape. We then project galaxies over a grid with approximately square pixels with side $0.85 \ {\rm arcmin}$ only considering systems with $0.1 \le z \le 1.4$ and using the sinusoidal 
projection to redefine the right ascension $\alpha$ as $(\alpha-\alpha_0) \cos{\delta}$.
The side of the pixel is chosen as a compromise between resolution and the need for a large enough number of galaxies in each pixel. Note that this is the same as the one used to build the convergence map from CFHTLenS data \citep{Ludo2013}.

The catalogs thus obtained are used as input for the estimate of high order convergence moments. We focus our attention on the the moments up to 4th order evaluated as 

\begin{displaymath}
\langle \kappa^n \rangle = \frac{1}{{\cal{N}}_{pix}} \sum{\tilde{\kappa}_i^n} \ \ (n = 2, 3, 4)
\end{displaymath}
where the sum is over the ${\cal{N}}_{pix}$ pixels in the map and $\tilde{\kappa}_i$ is the value of the convergence in the $i$\,-\,th pixel after smoothing the map with a given filter. Note that we first subtract a constant offset in order to have $\langle \kappa \rangle = 0$ although such a choice has no impact on the results. While $\langle \kappa^n \rangle$ is referred to as the moment of order $n$, it is useful to evaluate also two alternative quantities, namely the skewness $S_3$ and the kurtosis $S_4$ defined as

\begin{displaymath}
S_3 = \langle \kappa^3 \rangle/(\langle \kappa^2 \rangle)^{3/2} \ \ , \ \ 
S_4 = \langle \kappa^4 \rangle/(\langle \kappa^2 \rangle)^{2} \ \ , \ \ 
\end{displaymath}
which quantifies the deviation from a zero centred Gaussian distribution (having $S_3 = 0$ and $S_4 = 3$). 

An important preliminary step is the smoothing of the map which is tipically used to reduce the noise when working with actual noisy rather than simulated noiseless data. Following this common practice, we use three different filters, namely the Gaussian, top hat and aperture mass ones. In the Fourier space, the window functions read respectively

\begin{equation}
\tilde{W}(\ell, \theta_0) = \left \{
\begin{array}{l}
\displaystyle{\exp{(- \ell^2 \theta_0^2/2)}} \\
 \\ 
\displaystyle{2 J_1(\ell \theta_0)/(\ell \theta_0)} \\
 \\ 
\displaystyle{\sqrt{276} J_4(\ell \theta_0)/(\ell \theta_0)^2} \\
\end{array}
\right . 
\label{eq: filters}
\end{equation}
where $J_{\nu}$ is the Bessel function of order $\nu$ and $\theta_0$ the filter aperture which we vary over the range (2, 20) arcmin.

\section{Calibrating the moments}

High order moments are statistical quantities summarizing the properties of the convergence distribution. As can be easily understood, their values depend on which smoothing function is used and on the filter aperture for a given cosmological model. For a fixed filter, one can therefore use their dependence on $\theta_0$ as a tool to constrain cosmological parameters. To this end, one should impose 

\begin{equation}
\langle \kappa^n \rangle_{obs}(\theta_0) = \langle \kappa^n \rangle_{th}(\theta_0, {\bf p}_c)
\label{eq: obsvsthideal}
\end{equation}
with ${\bf p}_c$ the set of cosmological parameters. Actually, due to inaccuracies in the estimate of the observed moments $\langle \kappa^n \rangle_{obs}(\theta_0)$ and imperfect derivation\footnote{We refer the reader to Paper II for details.} of the theoretical moments $\langle \kappa^n \rangle_{th}(\theta_0)$, we do not expect the one\,-\,to\,-\,one relation (\ref{eq: obsvsthideal}) to actually hold. One can nevertheless still use moments to constrain the underlying cosmology provided a linear relation among theory and observed quantities hold. We therefore replace Eq.(\ref{eq: obsvsthideal}) with the more realistic one

\begin{equation}
\langle \kappa^n \rangle_{obs}(\theta_0) = (1 + m_n) \langle \kappa^n \rangle_{th}(\theta_0, {\bf p}_c) + c_n
\label{eq: obsvsth}
\end{equation}
with $(m_n, c_n)$ nuisance parameters introduced ad hoc. In order to make moments (but also skewness and kurtosis) valid tool for cosmology, we must show that this relation actually holds, investigate over which range it fits the data without introducing significant scatter, and check which is the minimum area needed to make the $(m, c)$ quantities stable. These are the topics we will investigate in the following. 

\begin{figure*}
\centering
\includegraphics[width=3.45cm]{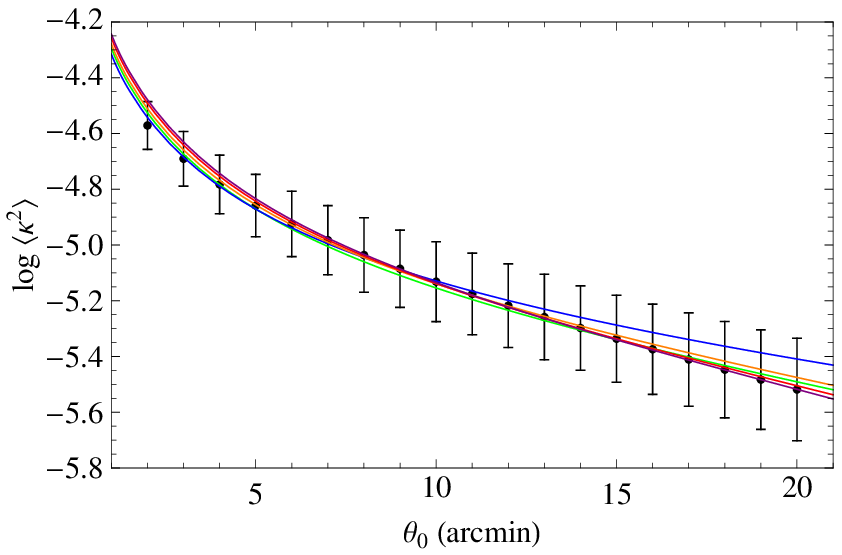}
\includegraphics[width=3.45cm]{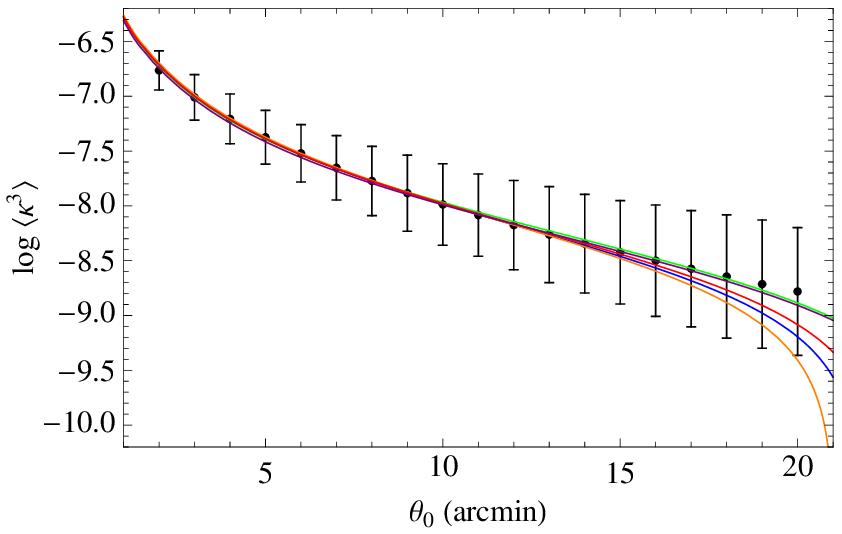}
\includegraphics[width=3.45cm]{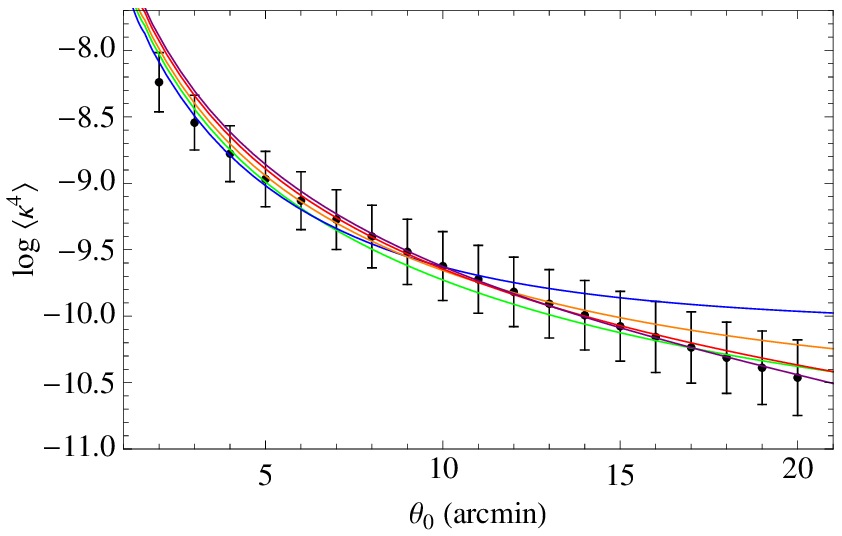} 
\includegraphics[width=3.45cm]{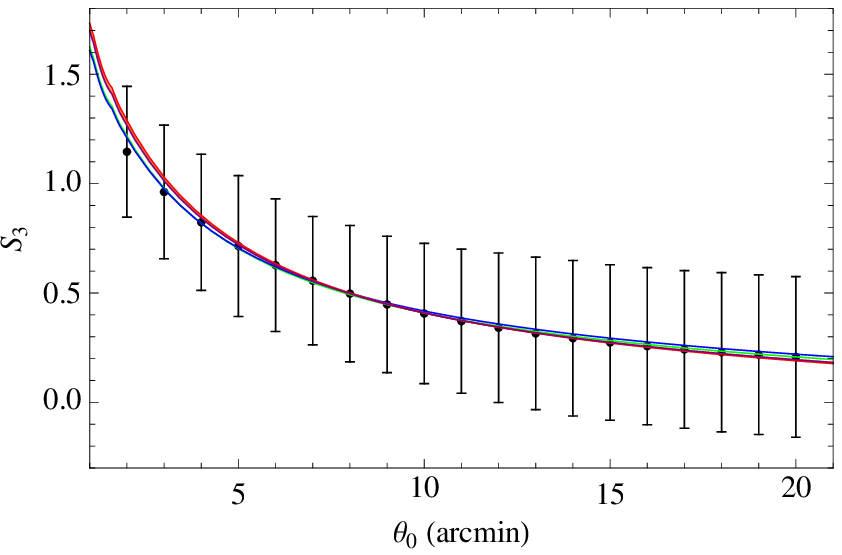} 
\includegraphics[width=3.45cm]{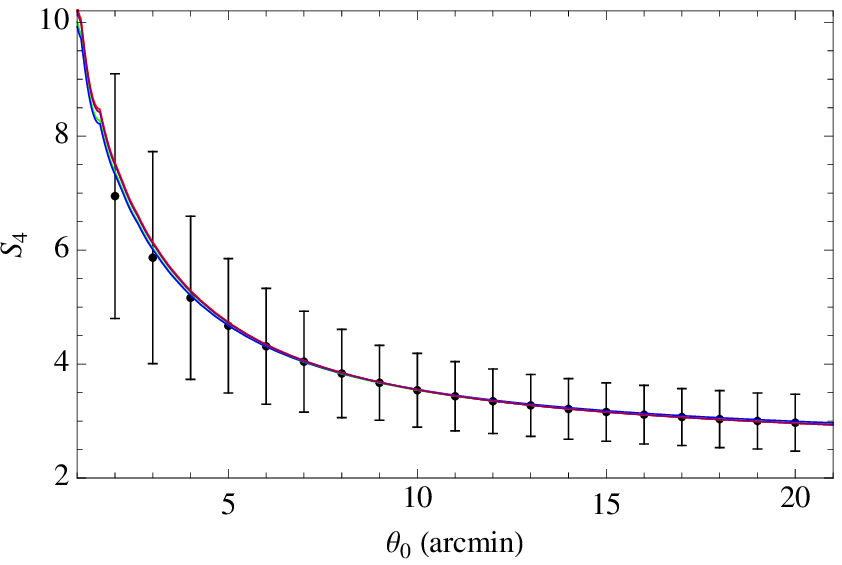}  \\
\includegraphics[width=3.45cm]{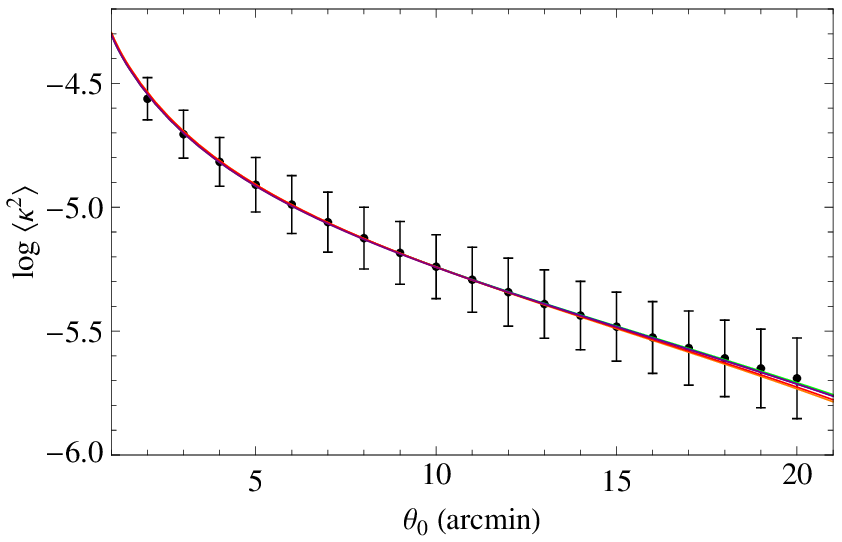}
\includegraphics[width=3.45cm]{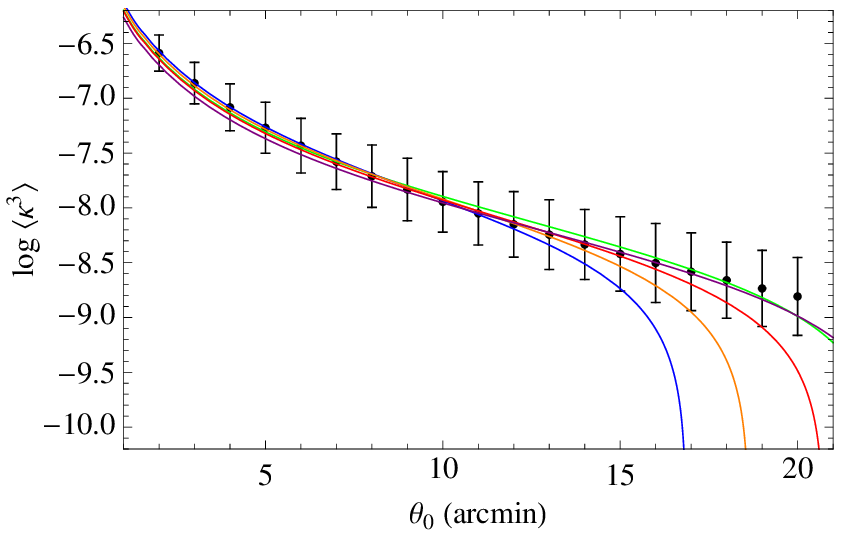}
\includegraphics[width=3.45cm]{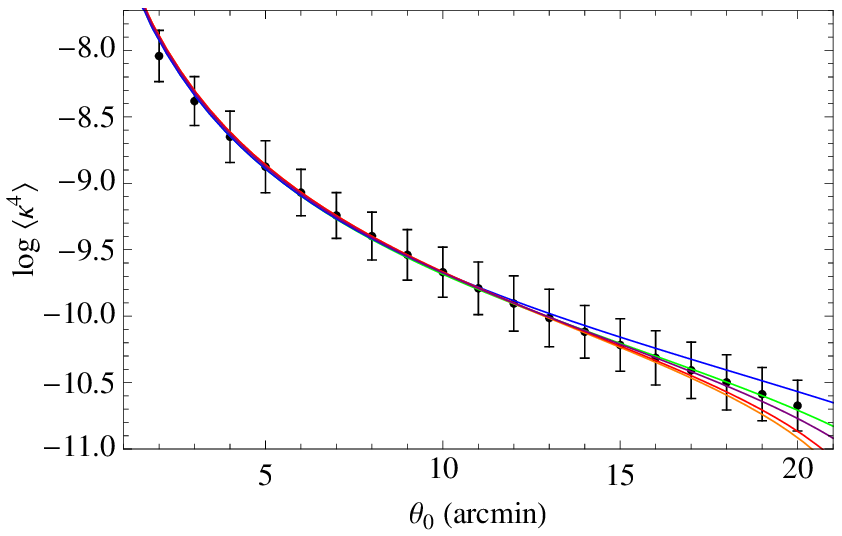} 
\includegraphics[width=3.45cm]{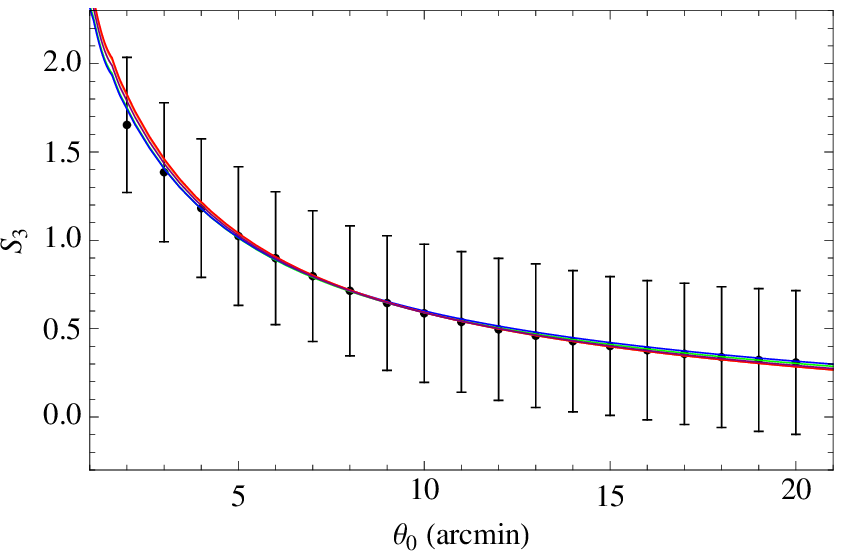} 
\includegraphics[width=3.45cm]{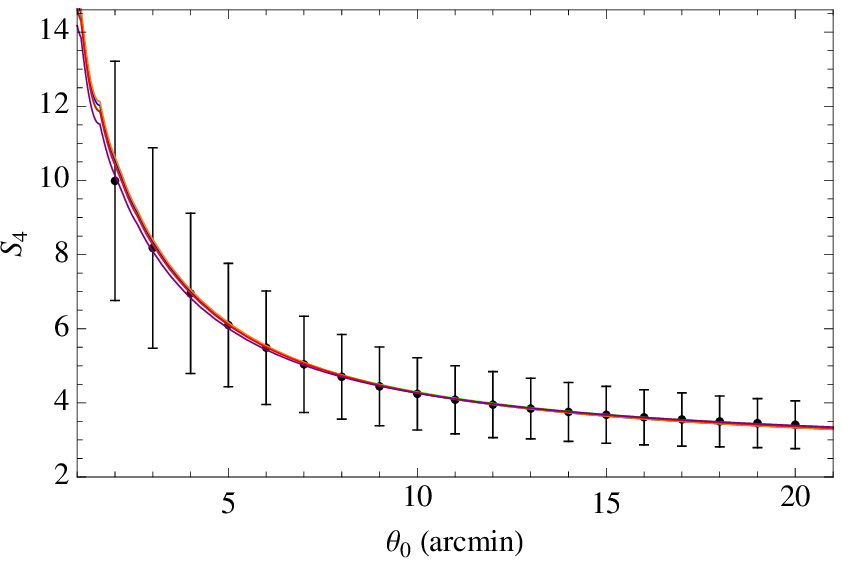} 
\caption{Moments, skewness and kurtosis for the Gaussian filter as measured on the simulated convergence map (top panels) and the reconstructed one (bottom panels) without noise and systematic errors. Green, blue, organge, red, purple curves refer to the linear relation (\ref{eq: obsvsth}) with $(m, c)$ from the best fit to the data over the ranges (2, 20), (2, 12), (4, 14), (6, 16), (8, 18) arcmin, respectively. Note that the curves are often superimposed so that discriminating among them is actually impossible.}
\label{fig: fitmomnonoise}
\end{figure*}

\begin{table*}
\begin{center}
\caption{Calibration parameters, $(m, c)$, and rms percentage residuals, $\rho_{rms} = rms[100 \times (1 - \mu_n(fit)/\mu_n(obs))]$, for different moments. Note that $c$ is in units of $(10^{-6}, 10^{-9}, 10^{-11})$ for $(\mu_2, \mu_3, \mu_4)$, respectively. A Gaussian filter is used to smooth the map.} 
\resizebox{17cm}{!}{
\begin{tabular}{cccccccccccccccc}
\hline \hline
\multicolumn{16}{c}{No mass reconstruction - No noise - No denoising} \\
\hline
$\theta_0$ range & \multicolumn{3}{c}{(2 - 20)} & \multicolumn{3}{c}{(2 - 12)} & \multicolumn{3}{c}{(4 - 14)} & \multicolumn{3}{c}{(6 - 16)} & \multicolumn{3}{c}{8 - 18} \\
\hline
Id & $m$ & $c$ & $\rho_{rms}$ & $m$ & $c$ & $\rho_{rms}$  & $m$ & $c$ & $\rho_{rms}$  & $m$ & $c$  & $\rho_{rms}$  & $m$ & $c$ & $\rho_{rms}$  \\
\hline
$\mu_2$ & $0.02 \pm 0.04$ & $-1.58 \pm 0.34$ & $4.24$ & 
$-0.05 \pm 0.05$ & $-0.59 \pm 0.58$ & $2.83$  & $0.05 \pm 0.03$ & $-1.63 \pm 0.32$ & $1.48$ & $0.11 \pm 0.03$ & $-2.15 \pm 0.18$ & $0.83$ & $0.15 \pm 0.01$ & $-2.43\pm 0.09$ & $0.44$ \\
$\mu_3$ & $0.51 \pm 0.05$ & $-3.07 \pm 0.29$ & $7.38$ & $0.56 \pm 0.06$ & $-3.88 \pm 0.81$ & $2.98$ & $0.61 \pm 0.03$ & $-4.29 \pm 0.31$ &$2.18$ & $0.55 \pm 0.06$ & $-3.66 \pm 0.41$ & $3.35$ & $0.43 \pm 0.07$  & $-2.92 \pm 0.37$ & $3.67$ \\
$\mu_4$ & $-0.10 \pm 0.12$ & $1.06 \pm 1.12$ & $18.4$ & $-0.22 \pm 0.12$ & $8.17 \pm 4.57$ & $13.74$ & $-0.00 \pm 0.09$ & $2.65 \pm 1.77$ & $7.20$ & $0.14 \pm 0.06$ & $0.33 \pm 0.75$ & $4.03$ & $0.24 \pm 0.04$ & $-0.70 \pm 0.33$ & $2.24$ \\
$S_3$ & $1.19 \pm 0.06$ & $-0.41 \pm 0.02$ & $2.35$ & $1.15 \pm 0.09$ & $-0.38 \pm 0.05$ & $2.54$ & $1.35 \pm 0.04$ & $-0.47 \pm 0.02$ & $0.63$ & $1.39 \pm 0.03$ & $-0.48 \pm 0.01$ & $0.60$ & $1.32 \pm 0.05$ & $-0.46 \pm 0.02$ & $0.97$ \\
$S_4$ & $-0.52 \pm 0.01$ & $2.24 \pm 0.02$ & $0.79$ & $-0.53 \pm 0.01$ & $2.27 \pm 0.05$ & $0.88$ & $-0.51 \pm 0.01$ & $2.20 \pm 0.01$ & $0.21$ & $-0.50 \pm 0.01$ & $2.19 \pm 0.01$ & $0.06$ & $-0.51 \pm 0.01$ & $2.20 \pm 0.01$ & $0.07$ \\
\hline \hline
\multicolumn{16}{c}{Yes mass reconstruction - No noise - No denoising} \\
\hline
$\theta_0$ range & \multicolumn{3}{c}{(2 - 20)} & \multicolumn{3}{c}{(2 - 12)} & \multicolumn{3}{c}{(4 - 14)} & \multicolumn{3}{c}{(6 - 16)} & \multicolumn{3}{c}{8 - 18} \\
\hline
Id & $m$ & $c$ & $\rho_{rms}$ & $m$ & $c$ & $\rho_{rms}$  & $m$ & $c$ & $\rho_{rms}$  & $m$ & $c$  & $\rho_{rms}$  & $m$ & $c$ & $\rho_{rms}$  \\
\hline
$\mu_2$ & $0.02 \pm 0.01$ & $-2.86 \pm 0.07$ & $1.19$ & $0.02 \pm 0.01$ & $-2.91 \pm 0.16$ & $0.86$ & $0.04 \pm 0.01$ & $-3.09 \pm 0.04$ & $0.21$ & $0.04 \pm 0.01$ & $-3.03 \pm 0.07$ & $0.43$ & $0.02 \pm 0.01$ & $-2.88 \pm 0.09$ & $0.64$ \\
$\mu_3$ & $0.86 \pm 0.17$ & $-4.37 \pm 0.75$ & $14.7$ & $1.19 \pm 0.08$ & $-8.48 \pm 0.90$ & $3.99$ & $1.03 \pm 0.13$ & $-6.65 \pm 1.03$ & $5.46$ & $0.82 \pm 0.13$ & $-4.96 \pm 0.82$ & $5.68$ & $0.60 \pm 0.12$ & $-3.62 \pm 0.57$ & $5.41$  \\
$\mu_4$ & $0.16 \pm 0.05$ & $-2.07 \pm 0.29$ & $6.17$ & $0.16 \pm 0.09$ & $-1.29 \pm 2.34$ & $7.22$ & $0.25 \pm 0.02$ & $-3.12 \pm 0.34$ & $1.53$ & $0.25 \pm 0.02$ & $-2.98 \pm 0.23$ & $1.54$ & $0.21 \pm 0.04$ & $-2.49 \pm 0.27$ & $2.41$ \\
$S_3$ & $2.14 \pm 0.06$ & $-0.58 \pm 0.03$ & $1.76$ & $2.10 \pm 0.12$ & $-0.56 \pm 0.06$ & $2.04$ & $2.33 \pm 0.04$ & $-0.65 \pm 0.02$ & $0.40$ & $2.35 \pm 0.04$ & $-0.66 \pm 0.01$ & $0.59$ & $2.25 \pm 0.07$ & $-0.62 \pm 0.03$ & $0.88$ \\
$S_4$ & $-0.23 \pm 0.01$ & $2.19 \pm 0.03$ & $0.75$ & $-0.22 \pm 0.02$ & $2.13 \pm 0.06$ & $0.75$ & $-0.21 \pm 0.01$ & $2.11 \pm 0.04$ & $0.38$ & $-0.23 \pm 0.02$ & $2.18 \pm 0.05$ & $0.39$ & $-0.27 \pm 0.02$ & $2.26 \pm 0.04$ & $0.26$ \\
\hline \hline
\end{tabular}}
\label{tab: mcvalsnonoise}
\end{center}
\end{table*}

As a preliminary remark, let us stress that $(m, c)$ are not forced to be small quantities. As will be detailed in Paper II, the theoretical moments can be estimated as 

\begin{equation}
\langle \kappa^2 \rangle_{th}(\theta_0) = {\cal{C}}_2(\kappa_{\theta_0}) \ ,
\label{eq: kappa2nd}
\end{equation}

\begin{equation}
\langle \kappa^3 \rangle_{th}(\theta_0) = 3 {\cal{Q}}_3 {\cal{C}}_3(\kappa_{\theta_0}^2) \ ,
\label{eq: kappa3rd}
\end{equation}

\begin{equation}
\langle \kappa^4 \rangle_{th}(\theta_0) = (12 {\cal{R}}_a + 4 {\cal{R}}_b) {\cal{C}}_4(\kappa_{\theta_0}^3) \ ,
\label{eq: kappa4th}
\end{equation}
where we have defined

\begin{equation}
\kappa_{\theta_0} = 2\pi \int{P[\ell/\chi(z), z] {\cal{W}}(\ell \theta_0) \ell d\ell} \ ,
\label{eq: defkappatheta0}
\end{equation}

\begin{equation}
{\cal{C}}_t(\kappa^n_{\theta_0}) = \frac{c}{H_0} \int_{0}^{z_h}{\frac{W^t(z) \kappa_{\theta_0}^{n}(z)}{\chi^{2(t - 1)}(z) E(z)} dz} \ .
\label{eq: defcfun}
\end{equation}
In Eq.(\ref{eq: defkappatheta0}), $P(k, z)$ is the matter power spectrum evaluated in $k = \ell/\chi(z)$ because of the flat sky approximation, while the coefficients $({\cal{Q}}_3, {\cal{R}}_a, {\cal{R}}_b)$ depends on the amplitude of the different tree topologies. Since different hierarchical models predict different values, we will fix them to some reference values, but allow for deviations as a consequence of our ignorance of the actual hierarchy. This motivates the introduction of a multiplicative bias $(1 + m_n)$ for 3rd and 4th order moments, but we retain it in the 2nd order moment too to allow for deviations from the theoretical formula because of noise and map reconstruction. Moreover, an additive bias $c_n$ can originate from cutting the integration range to avoid the highly nonlinear regime.

What is important to stress is that $(m_n, c_n)$ are not forced to be close to null quantities because of where they come from. However, we nevertheless expect that $1 + m$ is a positive quantity since, should this not be the case, one should deal with unphysically negative moments. While it is clear why $\langle \kappa^n \rangle$ with $n$ even number can not be negative, it is less immediate to grasp why this is the case for $n = 3$ and hence the skewness $S_3$. Actually, we expect the convergence distribution to deviate from the Gaussian one because of long tail to high $\kappa$ accounting for the presence of massive clusters. As such, the $\kappa$ distribution is positively skewed which makes $\langle \kappa^3 \rangle$ and $S_3$ positive quantities.

\subsection{Validating the linear calibration}

As a first test, we estimate moments from the convergence map as it is already available in the MICECATv2 catalog. Hereafter, to shorten the notation, we denote with $\mu_n(\theta_0, f)$ the moment $\langle \kappa^n \rangle$ after smoothing the map $f$ with a filter of aperture $\theta_0$. As final estimate, we consider the mean and variance of the distribution over the 140 fields we have selected as described above. Using this straightforward procedure, we derive the observed $\mu_n(\theta_0)$ values for $\theta_0$ running from 2 to 20 arcmin in steps of 1 arcmin, while the corresponding theoretical values are computed following \cite{MJ01} as detailed in Paper II. We then estimate $(m, c)$ and the rms of the percentage residuals $\rho_{rms}$ from the fit of Eq.(\ref{eq: obsvsth}) to the data thus obtained weighting each point with the inverse of the variance.

We first consider the ideal case when the true convergence map is available. Fitting Eq.(\ref{eq: obsvsth}) to the data thus obtained over different $\theta_0$ ranges, we get the results shown in top panel of Fig. \ref{fig: fitmomnonoise} and summarized in the upper half of Table \ref{tab: mcvalsnonoise} for the Gaussian filter\footnote{Hereafter, we will only discuss the results obtained adopting a Gaussian filter referring the reader to the tables in Appendix for the top hat and aperture mass cases. We, however, stress that, although the $(m, c)$ values are different, all the considerations we will do for the Gaussian case also hold for the top hat case, while the aperture mass filter gives worst results as we will discuss later.}. As it is apparent from both the plots and the values of $\rho_{rms}$, the linear relation works quite well in matching the observed and theoretical values with the only exception of the third order moment if the full range is used. Narrowing the 
 fitting range have a minor impact on the $(m, c)$ value which are roughly consistent, but allows to significantly reduce the rms scatter with the largest $\theta_0$ values providing better performances. This is expected given that the larger is the smoothing range, the more the large $\kappa$ tail of the convergence distribution is smoothed out. Since these values are related to nonlinear structures, the larger is $\theta_0$, the smaller is the contribution of nonlinearities which makes the linear approximation more efficient. 

A further lesson which can be learned from the $\rho_{rms}$ values is that skewness and kurtosis should be preferred over third and fourth order moments being the rms percentage residuals for $(S_3, S_4)$ definitely smaller than those for $(\mu_3, \mu_4)$. Again, this is expected. Indeed, the ratio entering the definition of $(S_3, S_4)$ partially cancels the uncertainties in theoretical modeling of moments thus working in favour of a first order correction as Eq.(\ref{eq: obsvsth}) indeed is.

The above results refer to a fully idealised situation. Even in absence of any noise, one actually has not direct access to the convergence $\kappa$, but rather to the shear components\footnote{Actually, what is observed are the two components $(e_1, e_2)$ of the ellipticity, but with no noise and intrisic alignment there is no difference with the shear components.} $(\gamma_1, \gamma_2)$. One can then reconstruct the $\kappa$ map using the \cite{KS93} method (hereafter, KS93). We first smooth the map using the chosen filter and then use the KS93 formalism to get the reconstructed convergence maps we use as input to the moments estimated and calibration procedure. Note that we have verified that fully consistent results are obtained if we first use the KS93 method and then smooth the convergence map (but see the hint in the conclusion about this point).

Bottom panels in Fig.\,\ref{fig: fitmomnonoise} and values in the lower half of Table \ref{tab: mcvalsnonoise} convincingly show that the reconstruction procedure has not affected the validity of the linear approximation. Although the $(m, c)$ parameters take different values, the rms of percentage residuals is still satisfactorily small. Somewhat suprisingly, $\rho_{rms}$ is actually smaller than for the true convergence map when the same moment and fitting range is assumed. It is hard to quantitatively explain such an outcome, but we can speculate that the KS93 method forces the reconstructed map to obey certain regularity properties which then propagates to the final $\mu_n(\theta_0)$ curves making them more similar to a linear function of the ideal theoretical model. Which underlying assumptions in the KS93 method are responsible for this unexpected behaviour is hard to say.

\begin{figure*}
\centering
\includegraphics[width=3.45cm]{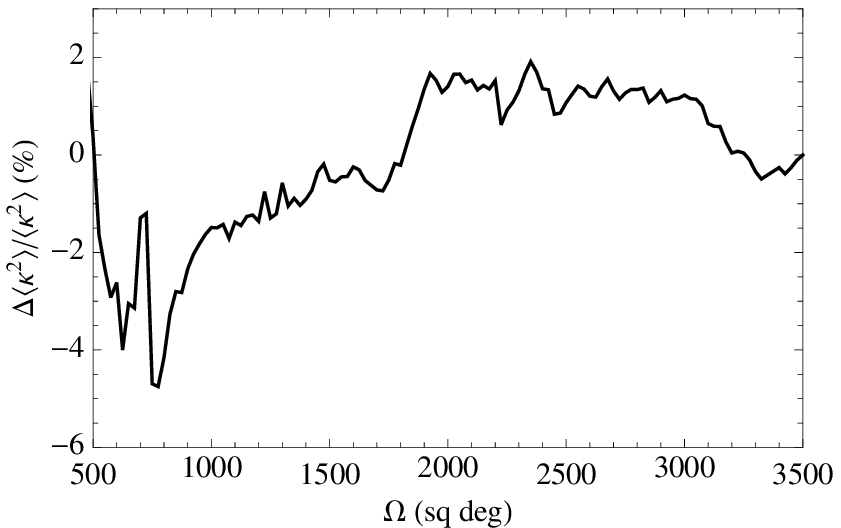}
\includegraphics[width=3.45cm]{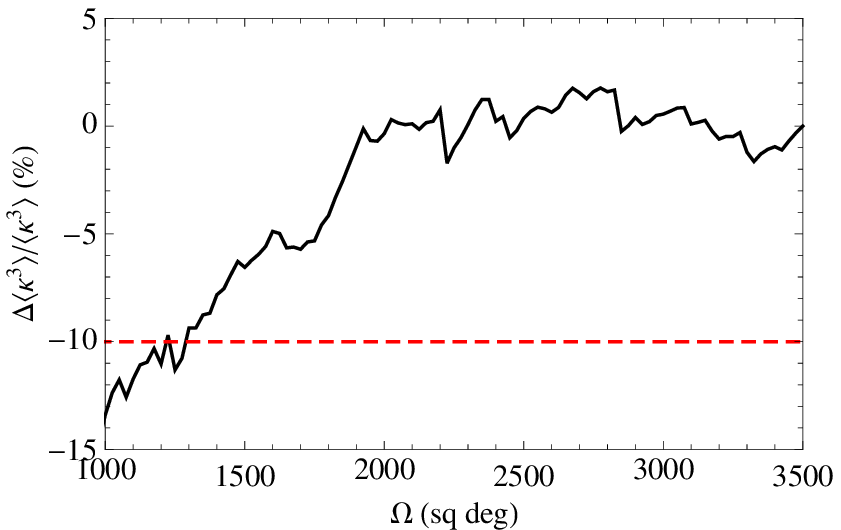}
\includegraphics[width=3.45cm]{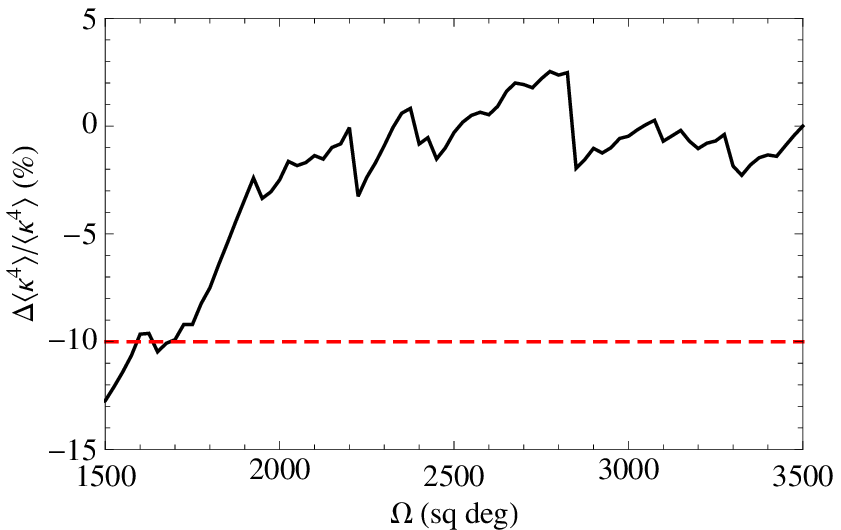} 
\includegraphics[width=3.45cm]{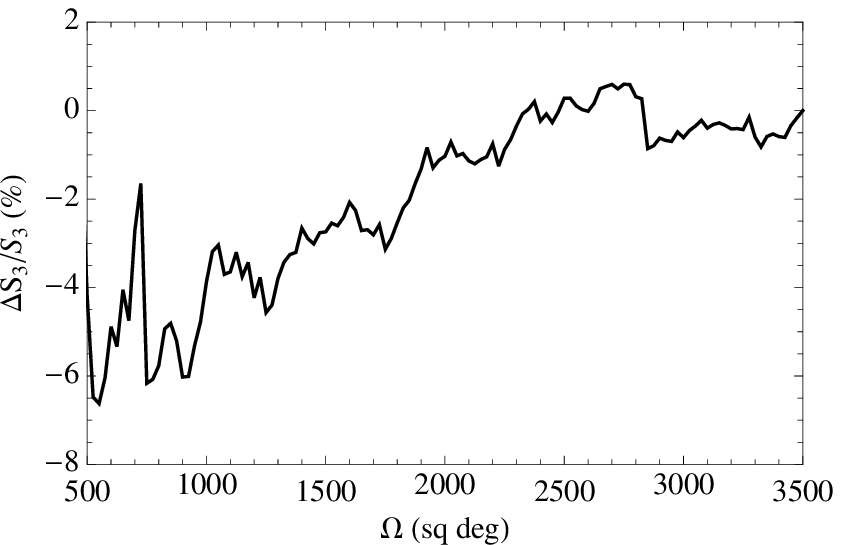} 
\includegraphics[width=3.45cm]{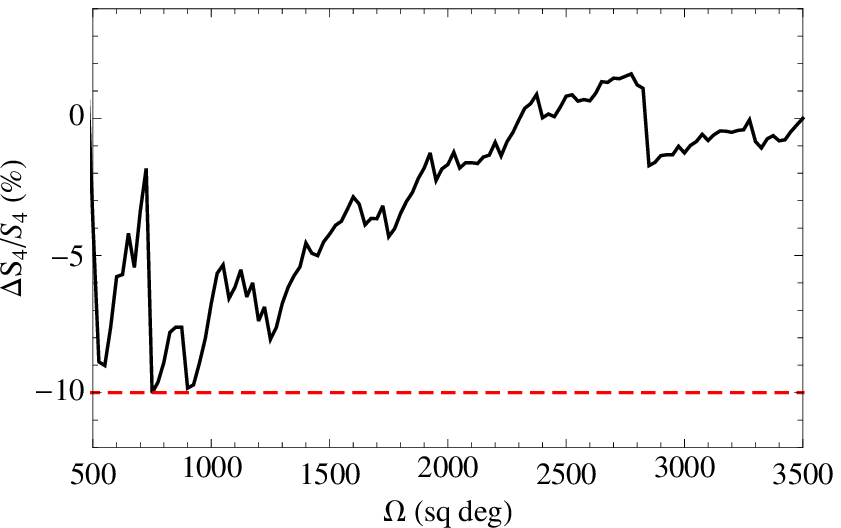}  \\
\includegraphics[width=3.45cm]{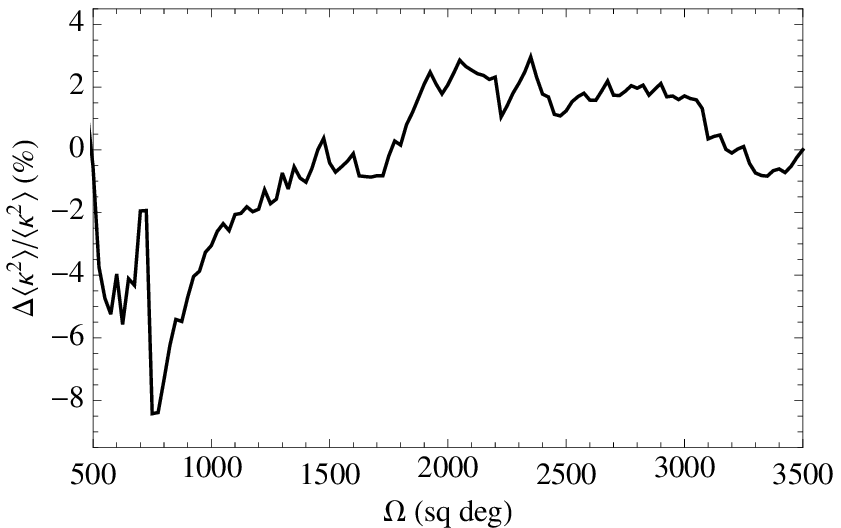}
\includegraphics[width=3.45cm]{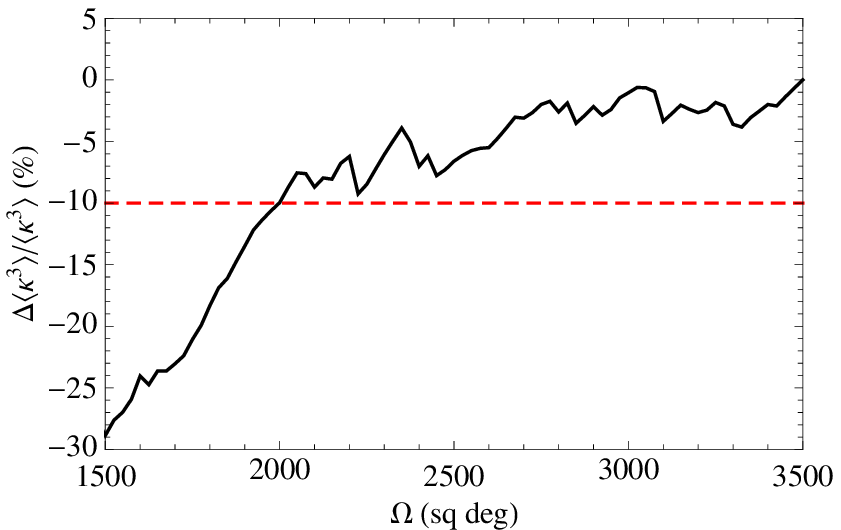}
\includegraphics[width=3.45cm]{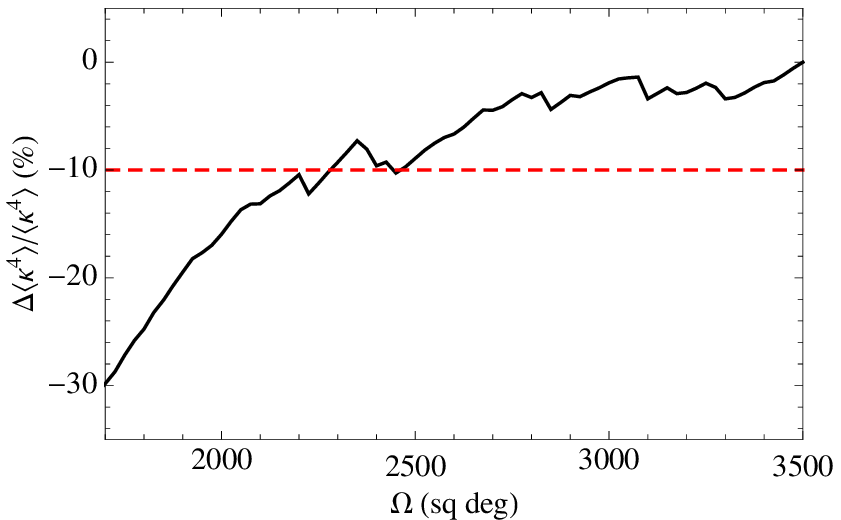} 
\includegraphics[width=3.45cm]{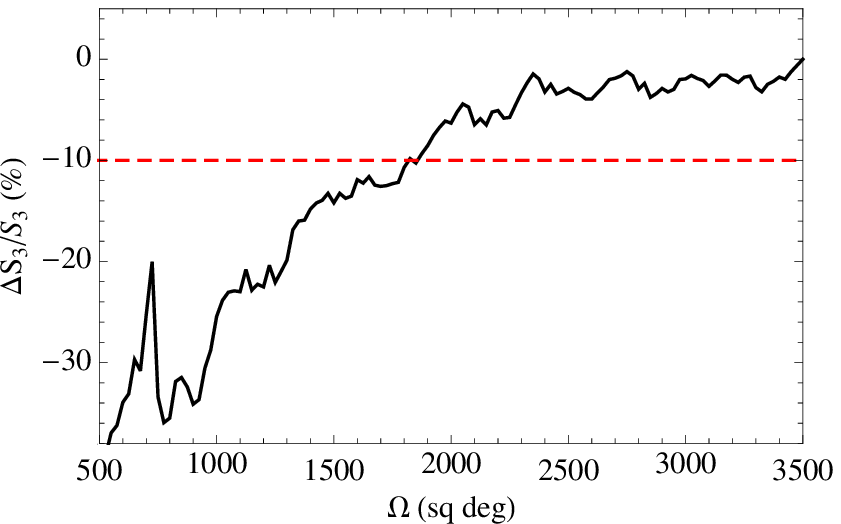} 
\includegraphics[width=3.45cm]{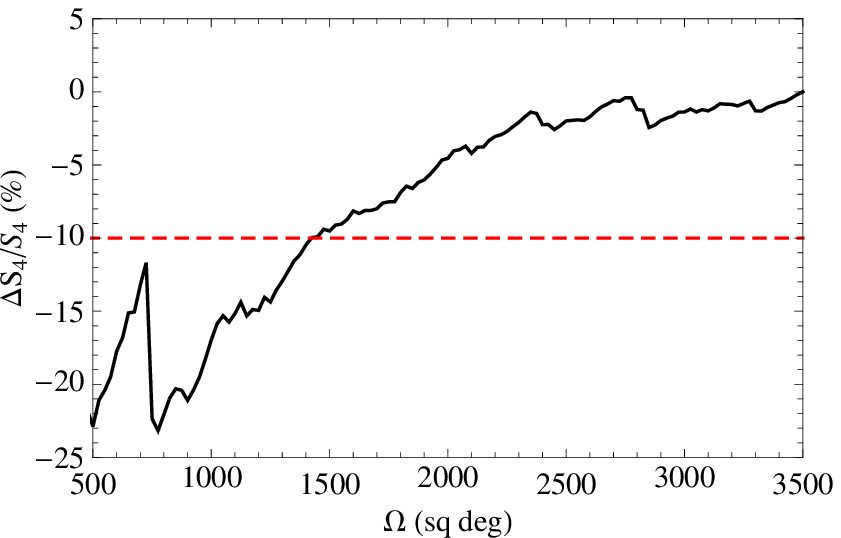}  \\
\includegraphics[width=3.45cm]{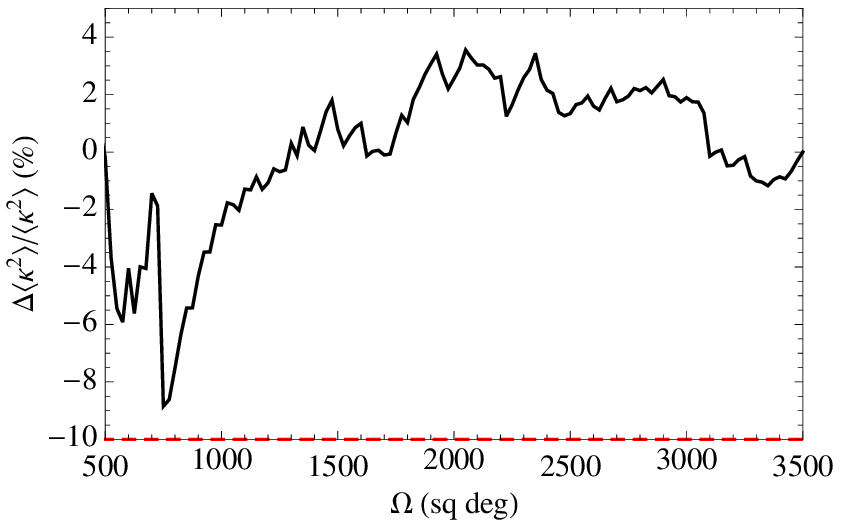}
\includegraphics[width=3.45cm]{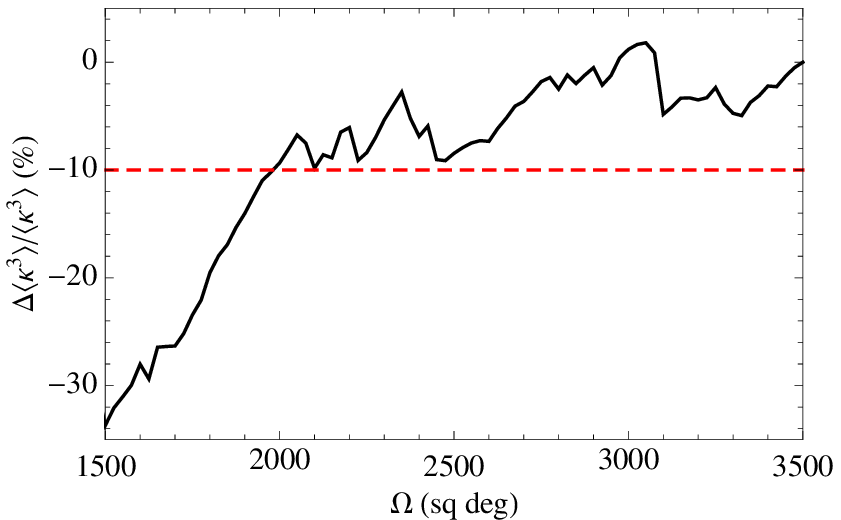}
\includegraphics[width=3.45cm]{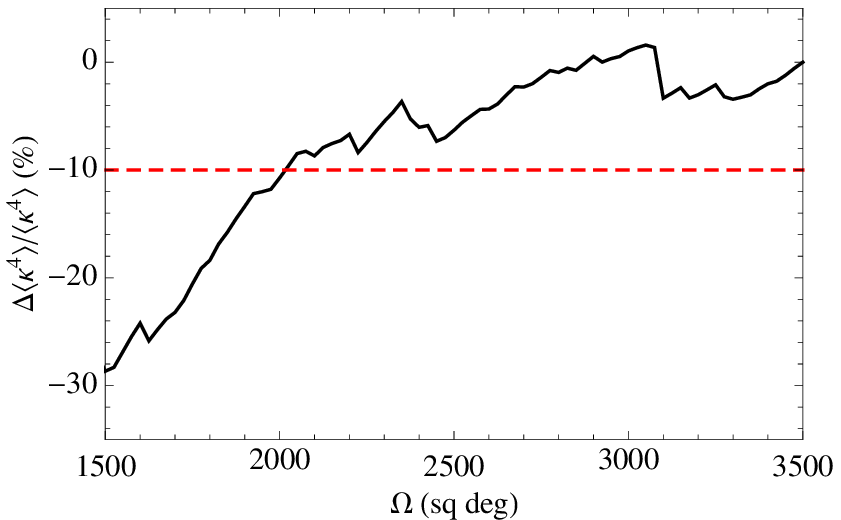} 
\includegraphics[width=3.45cm]{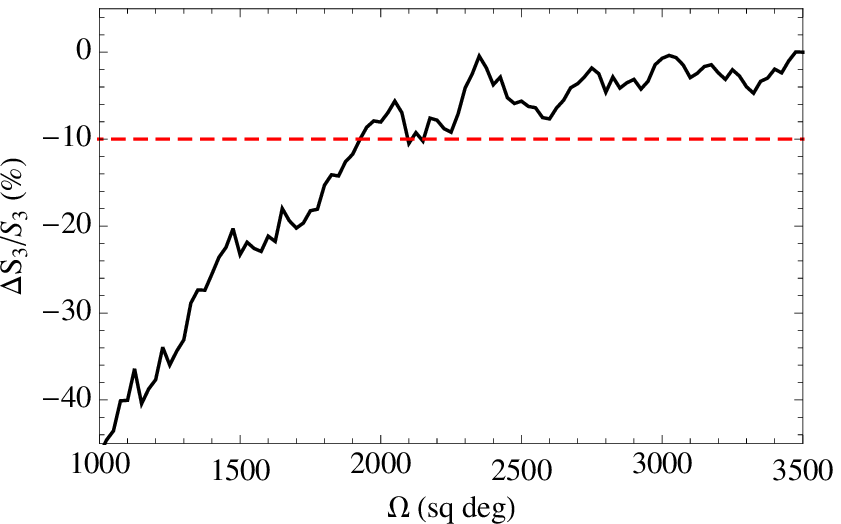} 
\includegraphics[width=3.45cm]{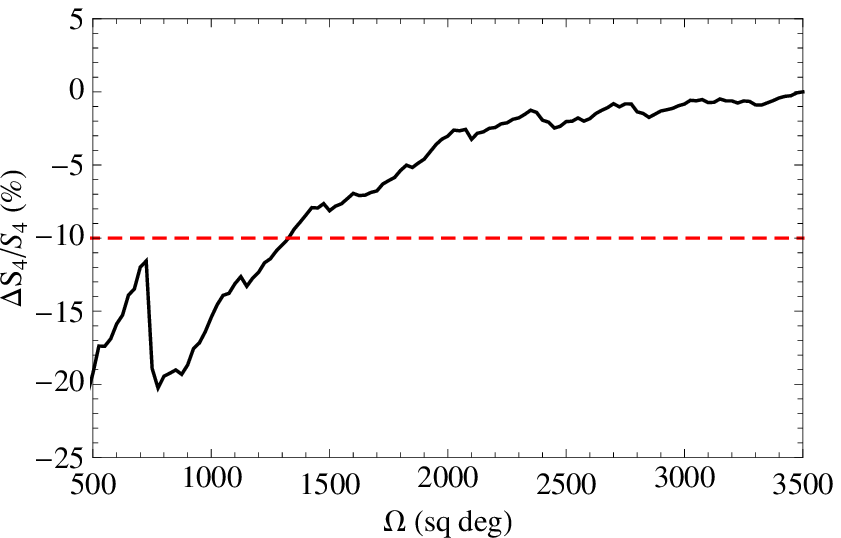}  \\
\caption{Percentage deviations of moments, skewness and kurtosis from the reference value for the Gaussian filter as measured on the reconstructed convergence map as function of the survey area for $\theta_0 = (2, 10, 18) \ {\rm arcmin}$ from top to bottom. Reference value is set to the one for $\Omega = 3500 \ {\rm sq \ deg}$. Beware of the different range on the $\Omega$\,-\,axis.}
\label{fig: convplot}
\end{figure*}

\begin{figure*}
\centering
\includegraphics[width=3.45cm]{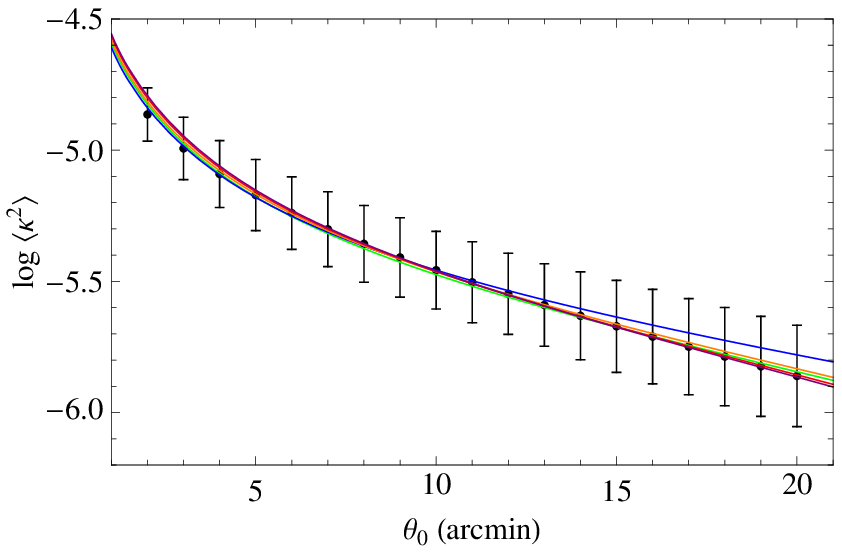}
\includegraphics[width=3.45cm]{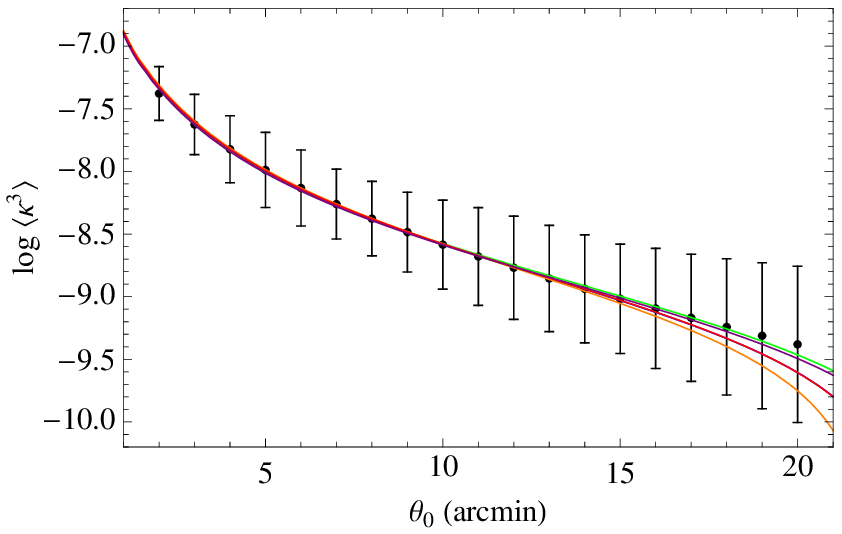}
\includegraphics[width=3.45cm]{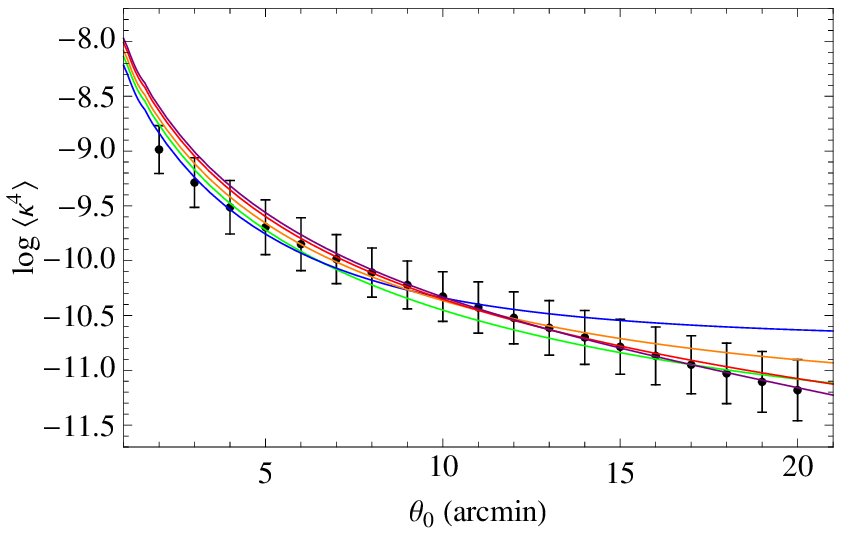} 
\includegraphics[width=3.45cm]{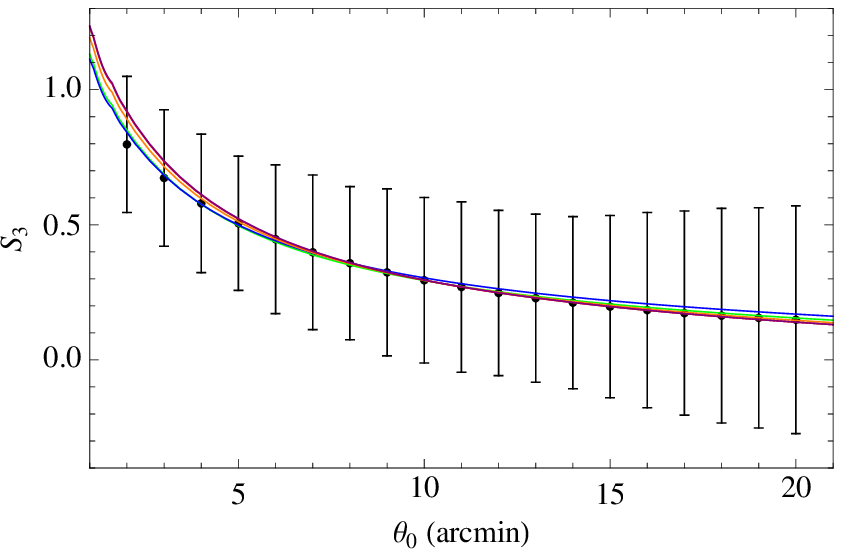} 
\includegraphics[width=3.45cm]{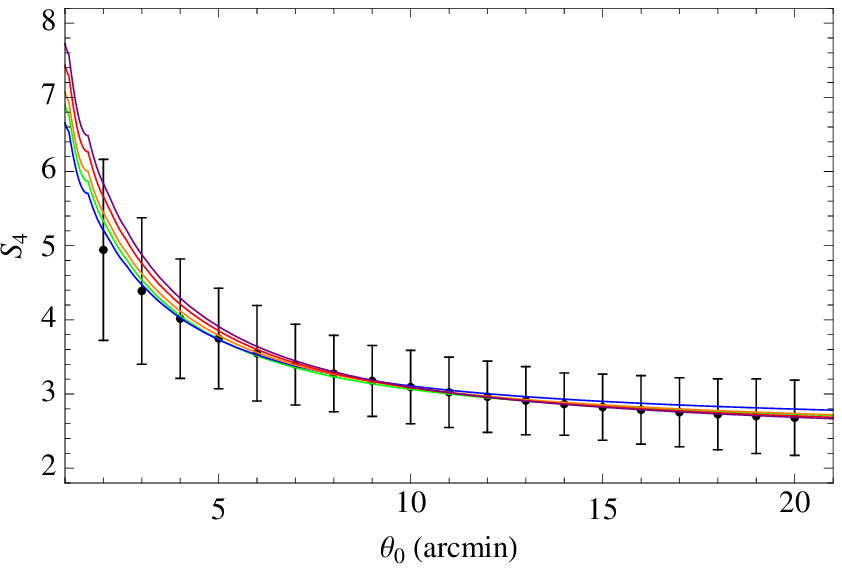}  \\
\includegraphics[width=3.45cm]{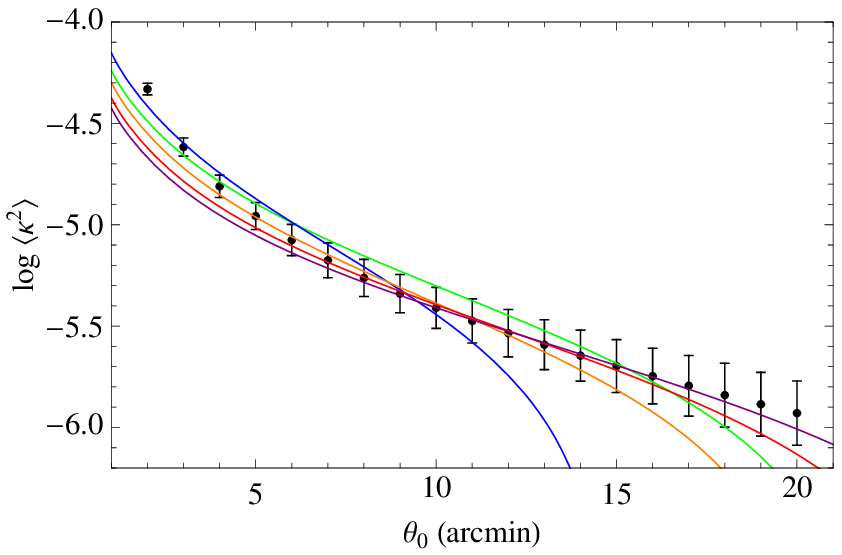}
\includegraphics[width=3.45cm]{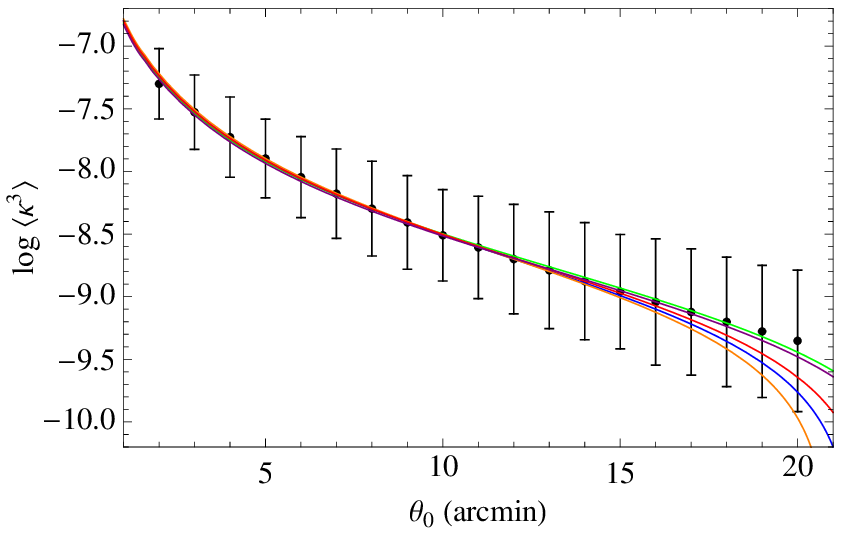}
\includegraphics[width=3.45cm]{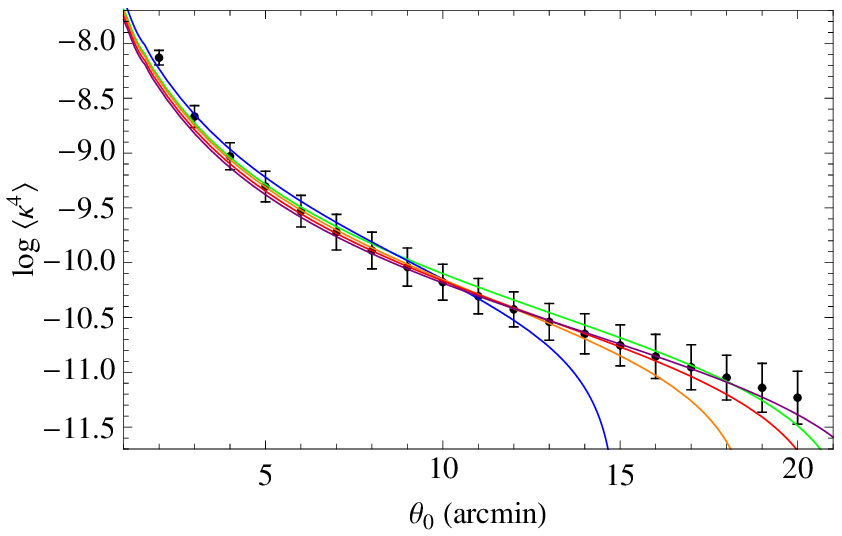} 
\includegraphics[width=3.45cm]{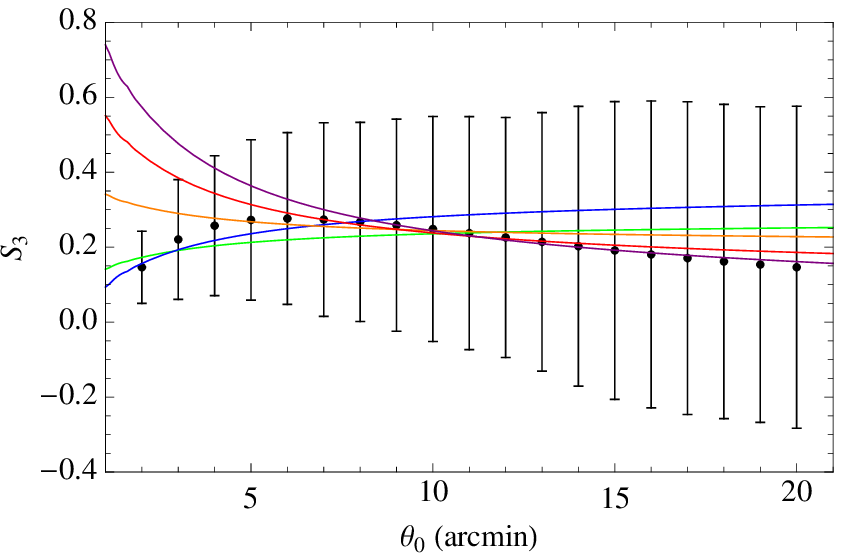} 
\includegraphics[width=3.45cm]{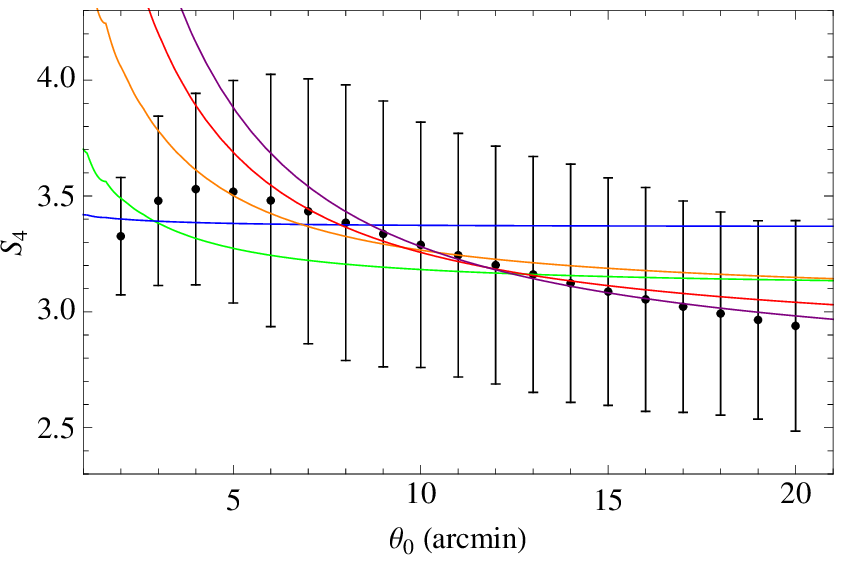} 
\caption{Same as Fig. \ref{fig: fitmomnonoise} but adding noise to the convergence (top panels) or the shear (bottom panels).}
\label{fig: fitmomyesnoise}
\end{figure*}

\begin{table*}
\begin{center}
\caption{Same as Table \ref{tab: mcvalsnonoise} but for noisy maps.} 
\resizebox{17cm}{!}{
\begin{tabular}{cccccccccccccccc}
\hline \hline
\multicolumn{16}{c}{No mass reconstruction - Yes noise - No denoising} \\
\hline
$\theta_0$ range & \multicolumn{3}{c}{(2 - 20)} & \multicolumn{3}{c}{(2 - 12)} & \multicolumn{3}{c}{(4 - 14)} & \multicolumn{3}{c}{(6 - 16)} & \multicolumn{3}{c}{8 - 18} \\
\hline
Id & $m$ & $c$ & $\rho_{rms}$ & $m$ & $c$ & $\rho_{rms}$  & $m$ & $c$ & $\rho_{rms}$  & $m$ & $c$  & $\rho_{rms}$  & $m$ & $c$ & $\rho_{rms}$  \\
\hline
$\mu_2$ & $-0.48 \pm 0.02$ & $-1.02 \pm 0.12$ & $3.02$ & $-0.51 \pm 0.02$ & $-0.67 \pm 0.21$ & $2.12$ & $-0.47 \pm 0.01$ & $-1.04 \pm 0.12$ & $1.17$ & $-0.45 \pm 0.01$ & $-1.22 \pm  0.06$ & $0.59$ & $-0.44 \pm 0.01$ & $-1.30 \pm 0.02$ & $0.24$ \\
$\mu_3$ & $-0.63 \pm 0.01$ & $-0.72 \pm 0.06$ & $5.65$ & $-0.63 \pm 0.01$ & $-0.85 \pm 0.17$ & $2.73$ & $-0.61 \pm 0.01$ & $-0.95 \pm 0.05$ & $1.28$ & $-0.62 \pm 0.01$ & $-0.85 \pm 0.07$ & $2.29$ & $-0.64 \pm 0.01$ & $-0.72 \pm 0.07$ & $2.97$ \\
$\mu_4$ & $-0.83 \pm 0.03$ & $0.25 \pm 0.22$ & $21.3$ & $-0.86 \pm 0.03$ & $1.86 \pm 0.91$ & $15.2$ & $-0.81 \pm 0.02$ & $0.58 \pm 0.41$ & $8.85$ & $-0.77 \pm 0.01$ & $0.06 \pm 0.18$ & $5.04$ & $-0.75 \pm 0.01$ & $-0.16 \pm 0.08$ & $2.63$ \\
$S_3$ & $0.51 \pm 0.05$ & $-0.27 \pm 0.02$ & $3.82$ & $0.46 \pm 0.07$ & $-0.24 \pm 0.04$ & $2.89$ & $0.62 \pm 0.04$ & $-0.31 \pm 0.02$ & $1.13$ & $0.69 \pm 0.02$ & $-0.34 \pm 0.01$ & $0.32$ & $0.69 \pm 0.02$ & $-0.34 \pm 0.01$ & $0.67$ \\
$S_4$ & $-0.71 \pm 0.02$ & $2.29 \pm 0.05$ & $1.67$ & $-0.74 \pm 0.02$ & $2.39 \pm 0.06$ & $1.19$ & $-0.71 \pm 0.01$ & $2.28 \pm 0.04$ & $0.61$ & $-0.68 \pm 0.01$ & $2.21 \pm 0.03$ & $0.37$ & $-0.66 \pm 0.01$ & $2.16 \pm 0.02$ & $0.21$ \\
\hline \hline
\multicolumn{16}{c}{Yes mass reconstruction - Yes noise - No denoising} \\
\hline
$\theta_0$ range & \multicolumn{3}{c}{(2 - 20)} & \multicolumn{3}{c}{(2 - 12)} & \multicolumn{3}{c}{(4 - 14)} & \multicolumn{3}{c}{(6 - 16)} & \multicolumn{3}{c}{8 - 18} \\
\hline
Id & $m$ & $c$ & $\rho_{rms}$ & $m$ & $c$ & $\rho_{rms}$  & $m$ & $c$ & $\rho_{rms}$  & $m$ & $c$  & $\rho_{rms}$  & $m$ & $c$ & $\rho_{rms}$  \\
\hline
$\mu_2$ & $0.22 \pm 0.21$ & $-5.32 \pm 1.49$ & $25.6$ & $0.55 \pm 0.30$ & $-9.50 \pm 3.23$ & $18.9$ & $0.07 \pm 0.11$ & $-4.96 \pm 0.96$ & $7.24$ & $-0.11 \pm 0.06$ & $-3.46 \pm  0.48$ & $4.19$ & $-0.22 \pm 0.04$ & $-2.72 \pm 0.29$ & $3.03$ \\
$\mu_3$ & $-0.55 \pm 0.02$ & $-0.94 \pm 0.08$ & $6.76$ & $-0.53 \pm 0.02$ & $-1.18 \pm 0.27$ & $3.70$ & $-0.52 \pm 0.01$ & $-1.29 \pm 0.09$ & $1.88$ & $-0.54 \pm 0.01$ & $-1.11 \pm 0.09$ & $2.63$ & $-0.57 \pm 0.02$ & $ -0.93 \pm  0.09$  & $3.19$ \\
$\mu_4$ & $-0.53 \pm 0.07$ & $-1.31 \pm 0.48$ & $20.3$ & $-0.43 \pm 0.11$ & $-4.13 \pm 2.51$ & $16.8$ & $-0.55 \pm 0.03$ & $-1.81 \pm  0.43$ & $5.67$ & $-0.59 \pm 0.02$ & $-1.23 \pm 0.21$ & $4.11$ & $-0.62 \pm 0.02$ & $-0.90 \pm 0.13$ & $4.03$ \\
$S_3$ & $-1.17 \pm 0.14$ & $0.30 \pm 0.08$ & $31.9$ & $-1.33 \pm 0.15$ & $0.41 \pm 0.09$ & $14.1$ & $-0.82 \pm 0.16$ & $0.18 \pm 0.07$ & $7.54$ & $-0.43 \pm 0.15$ & $0.03 \pm 0.06$ & $5.08$ & $-0.10 \pm 0.13$ & $-0.09 \pm 0.05$ & $3.23$ \\
$S_4$ & $-0.96 \pm 0.02$ & $3.08 \pm 0.13$ & $4.61$ & $-1.00 \pm 0.02$ & $3.36 \pm 0.17$ & $3.12$ & $-0.90 \pm 0.03$ & $2.99 \pm 0.11$ & $1.55$ & $-0.82 \pm 0.03$ & $2.76 \pm 0.08$ & $0.90$ & $-0.75 \pm 0.03$ & $2.59 \pm 0.06$ & $0.59$ \\
\hline \hline
\end{tabular}}
\label{tab: mcvalsyesnoise}
\end{center}
\end{table*}

\subsection{Survey area requirements}

The above results have been obtained using the full 3500 sq deg cut from the MICECATv2 catalog. While such a large area is well within the range of ongoing and future lensing surveys, it is nevertheless worth wondering which is the minimum area needed to get a reliable estimate of the moments. To this end, we plot in Fig. \ref{fig: convplot} the quantity

\begin{displaymath}
\Delta y/y = 100 \times [1 - y(\theta_0, \Omega)/y(\theta_0, \Omega = 3500 \ {\rm sq \ deg})]
\end{displaymath}
with $y = \mu_2, \mu_3, \mu_4, S_3, S_4$. Imposing $|\Delta y/y| < \Delta_{max}$, we can infer which is the minimum area a survey should have to be sure that the estimated moments are stable against cosmic variance. The value of $\Delta_{max}$ should be set according to the typical uncertainty and the quantity $y$ which is considered thus also taking into account the filter adopted. In order to set $\Delta_{max}$. we assume a Gaussian filter and take the reconstructed map as input for the estimate of the moments. The median uncertainty on $(\mu_2, \mu_3, \mu_4, S_3, S_4)$ may change by a factor of 2 from one $\theta_0$ value to another, but it is never smaller than $\sim 20\%$, while it can get as large as $100\%$ at large $\theta_0$. As a conservative choice, we therefore set $\Delta_{max} = 10\%$ and ask that $|\Delta y/y| < 10\%$ for all the moments and at all smoothing scales. 

As can be read from Fig. \ref{fig: convplot}, a survey area as large as $\Omega \simeq 2000 \ {\rm sq \ deg}$ is enough to fulfill the requirements on the stability of $(\mu_2, \mu_3, \mu_4, S_3, S_4)$. Note that this constraint is actally over conservative. First, we are including in the requirement also smoothing lenghts which will likely not be used in final cosmological analysis. Second, we are asking that the condition quoted above holds for $(\mu_3, \mu_4)$ and $(S_3, S_4)$. In practical applications, one will use the skewness and kurtosis only since they are less biased and can be estimated with a smaller error. Should we relax the requiremnt including only $(\mu_2, S_3, S_4)$ and smoothing lengths smaller than $\theta_0 = 14 \ {\rm arcmin}$, the mimimal survey area would decrease to $\Omega \simeq 1000 \ {\rm sq \ deg}$. Finally, we remind that we are using here the reconstructed map from noiseless data. Including noise will increase the errors on the measured moments
  thus allowing to accept larger $\Delta_{max}$ values and hence a smaller minimal area. We nevertheless prefer to be conservative and conclude that, in order to get a stable estimate of the moments, the survey area should be larger than $2000 \ {\rm sq \ deg}$.

\section{Moments from noisy maps}

The analysis in the previous section relies on noiseless data. Needless to say, this is far from be true in realistic applications. We therefore investigate here the impact of noise on the moments estimate and the validity of the linear relation between observed and theoretical quantities.

As a first step, we add noise directly to the convergence map. Following \cite{Hama2004}, we add to each pixel a random noise extracted from a normal distribution with zero mean and dispersion

\begin{equation}
\sigma_{pix}^2 = \frac{\sigma_\epsilon^2}{2}\frac{1}{\theta_{pixel}^2 n_g}
\end{equation}
where $\sigma_\epsilon$ is the variance of the intrinsic ellipticity distribution and $n_g$ the source number density. We set $\sigma_\epsilon = 0.28$, while it is $n_g = 26 \ {\rm gal/arcmin^2}$ for the MICECATv2 sample (averaging over the 140 fields we have cutted). Since noise is a random quantity, we estimate $(\mu_2, \mu_3, \mu_4, S_3, S_4)$ from 100 realizations containing the same signal (i.e., the input convergence map) but different noise. Moments are finally estimated averaging the results of the 100 noisy maps and are shown in Fig. \ref{fig: fitmomyesnoise}, while the $(m, c)$ coefficients  and the rms percentage residuals are summarized in the upper half of Table \ref{tab: mcvalsyesnoise} for the same fitting ranges of the noiseless case.

A comparison of corresponding panels in Fig. \ref{fig: fitmomnonoise} and \ref{fig: fitmomyesnoise} clearly shows the strong impact of noise on the estimated moments. It is worth discussing each order separetely. Let us first consider $\mu_2$. Although the shape of the $\mu_2(\theta_0)$ curve is roughly the same, its amplitude is decreased. Although this is not straightforward to appreciate from the logarithmic plots, we indeed get

\begin{displaymath}
\overline{\mu_2(\theta_0, {\rm noise}) /\mu_2(\theta_0, {\rm no \ noise})} = 0.47 \pm 0.01
\end{displaymath}
where we use the notation $\overline{x}$ $(\langle x \rangle)$ to denote the median (mean) value of $x$ and the error is the median deviation. Such a decrease is somewhat counterintuitive, but it is actually a consequence of how noisy moments have been estimated. Indeed, looking at the 100 realizations, it is possible to see that $\mu_2$ can be both larger and smaller than in  the noiseless case. In particular, for some realizations, one can also get $\mu_2 < 0$ so that averaging over both positive and negative values pushes the mean value to a final result smaller than in the noiseless case. As a consequence, the linear matching with the theoretical values given by Eq.(\ref{eq: obsvsth}) is only possible if $m < 0$ so that $1 + m < 1$ and $\mu_2$ is decreased. Note that, in realistic applications, one has a single unknown noise realization so that the result presented here is hard to extrapolate to actual data. However, we remind the reader that one would never deal with the
  convergence map itself, but rather with its reconstructed version so that we consider the case we are discussing only as a {\it gedancken experiment} useful to highlight the impact of noise. 

The consequences of adding noise can be qualitatively understood approximating the convergence distribution as a Gaussian one (which is not because of the high $\kappa$ tail due to massive clusters and nonlinearities). In this case, the combination of $\kappa$ and a Gaussian noise leads to a distribution which is more and more Gaussian as the noise dominates. Should this be the case, both the third order moment and the skewness would be smaller than in the noiseless case. Indeed, we get

\begin{displaymath}
\overline{\mu_3(\theta_0, {\rm noise}) /\mu_3(\theta_0, {\rm no \ noise})} = 0.252 \pm 0.003 \ ,
\end{displaymath}

\begin{displaymath}
\overline{S_3(\theta_0, {\rm noise}) /S_3(\theta_0, {\rm no \ noise})} = 0.716 \pm 0.007 \ ,
\end{displaymath}
which goes in the right direction. Note that the effect is smaller on the skewness because of the presence of the $\mu_2$ term in the definition which does not change in the expected direction because of what we have explained before. The same argument above also drives the change in the fourth order moment and the kurtosis. We find

 \begin{figure*}
\centering
\includegraphics[width=3.45cm]{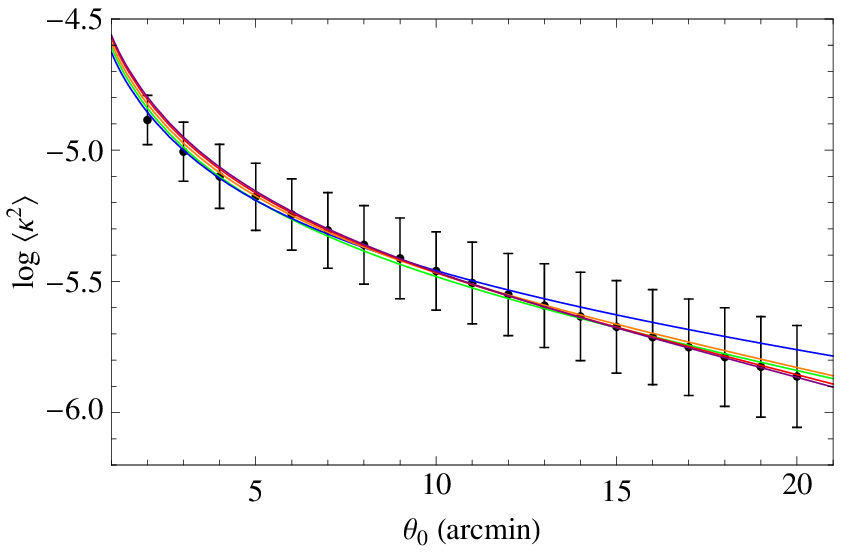}
\includegraphics[width=3.45cm]{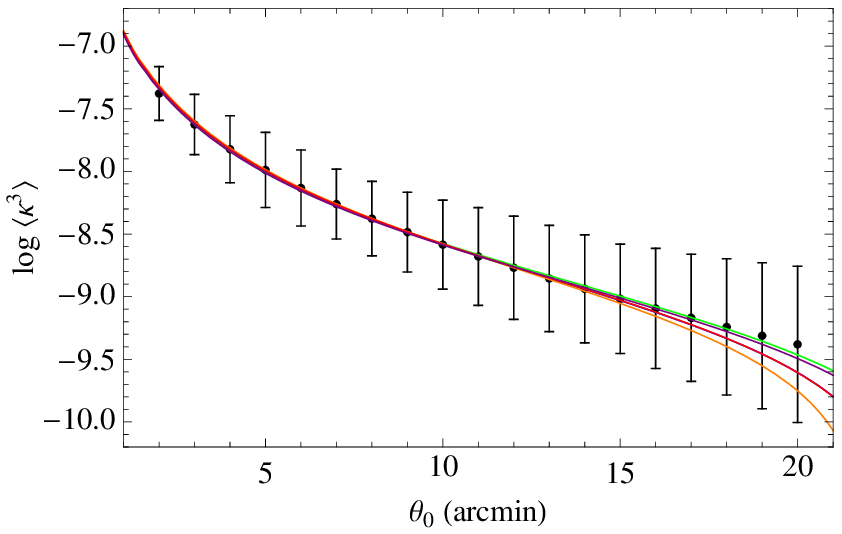}
\includegraphics[width=3.45cm]{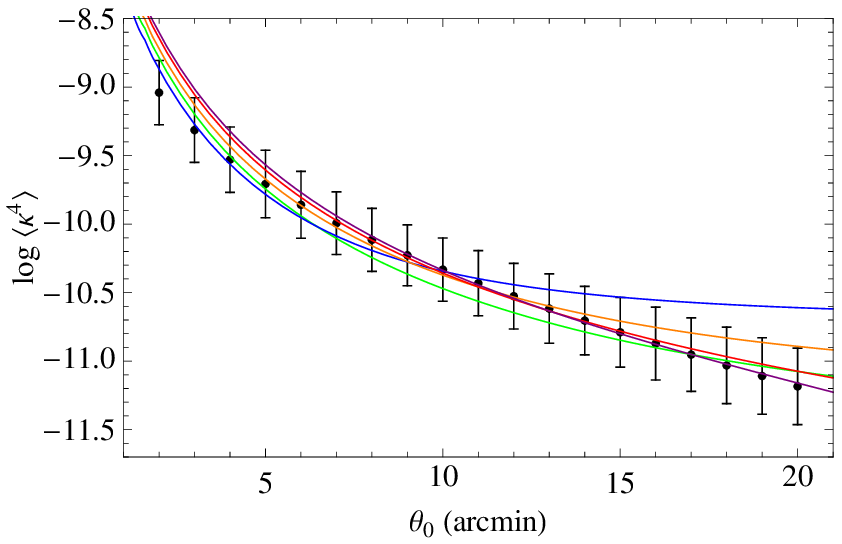} 
\includegraphics[width=3.45cm]{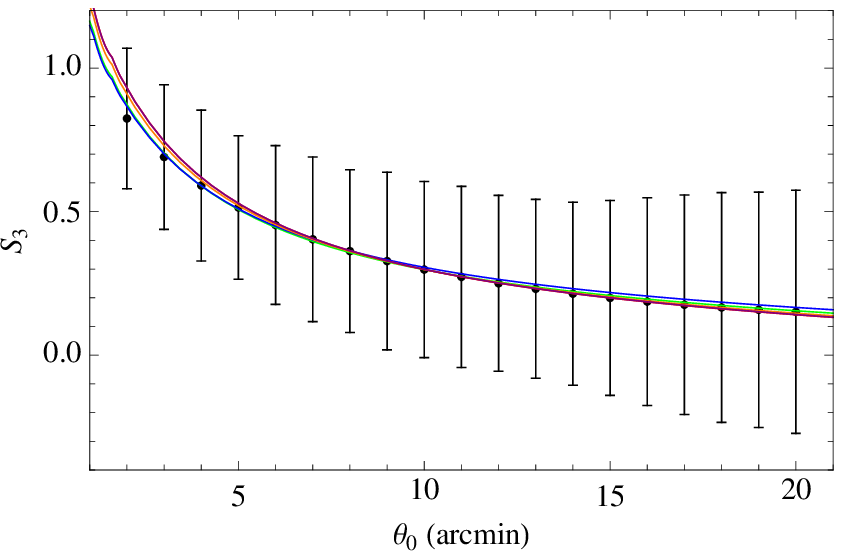} 
\includegraphics[width=3.45cm]{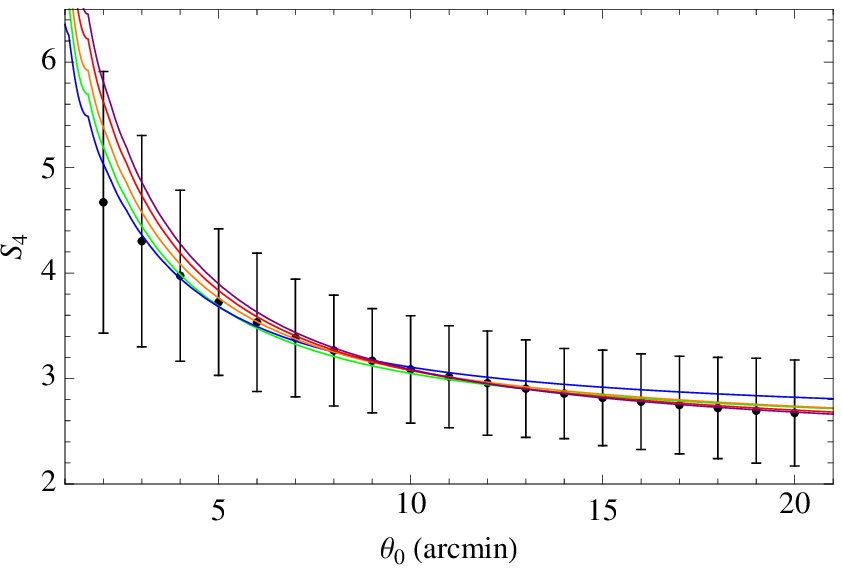}  \\
\includegraphics[width=3.45cm]{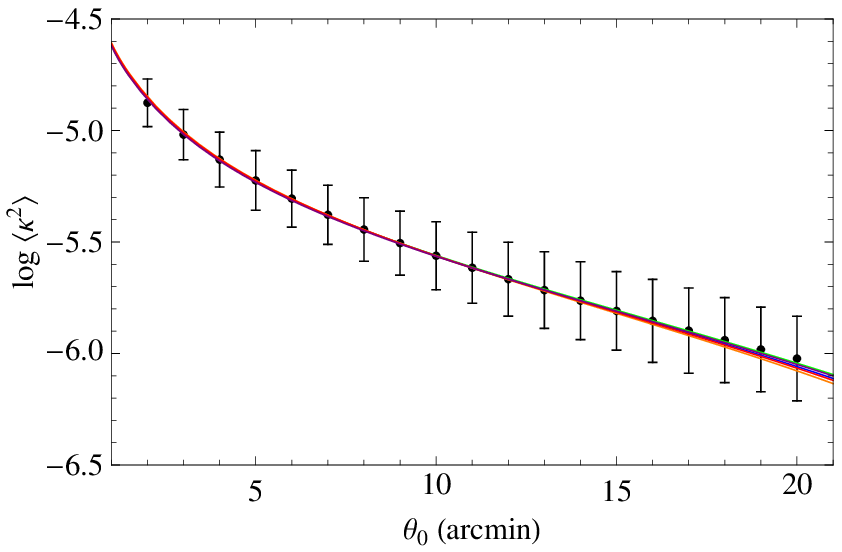}
\includegraphics[width=3.45cm]{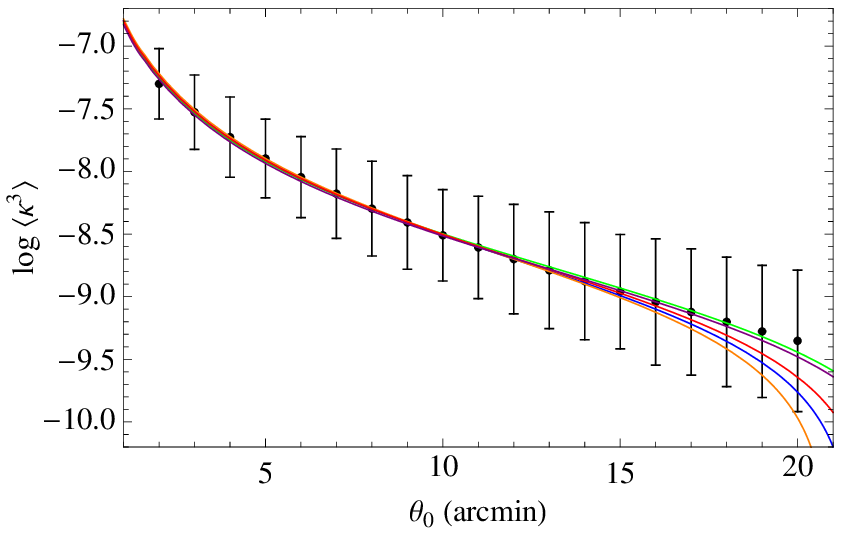}
\includegraphics[width=3.45cm]{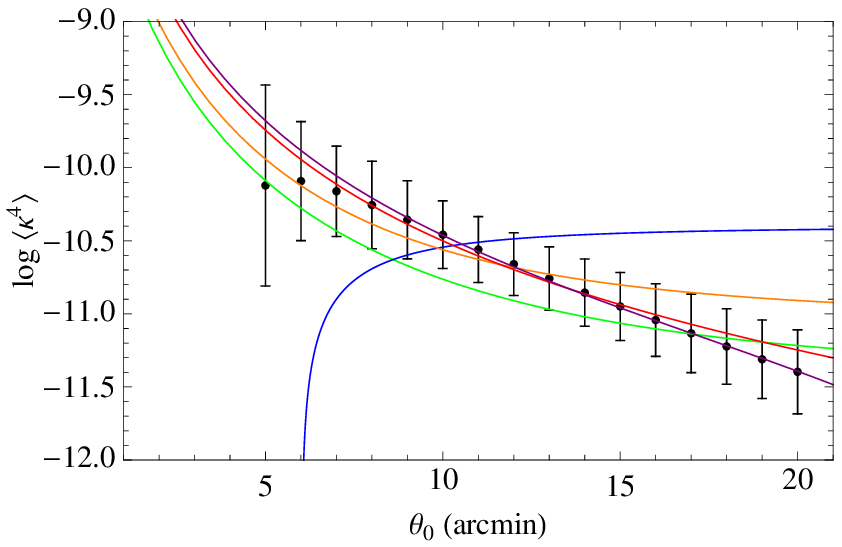} 
\includegraphics[width=3.45cm]{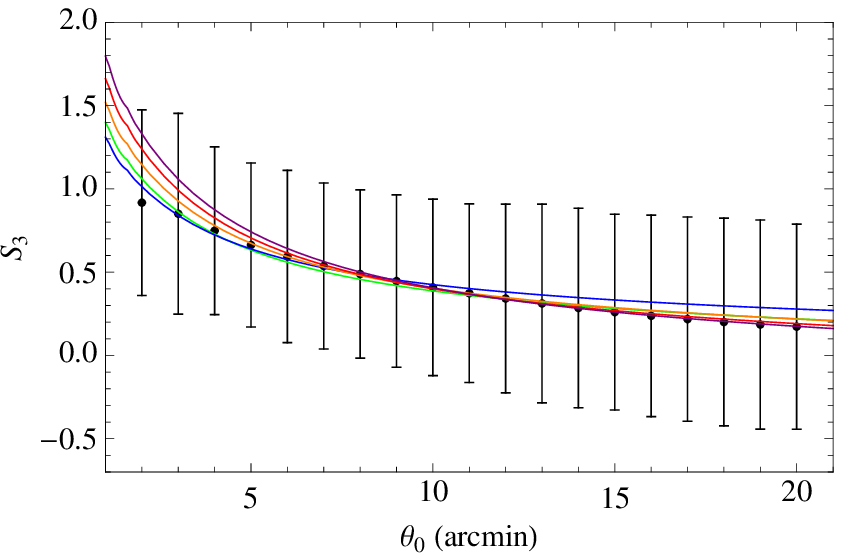} 
\includegraphics[width=3.45cm]{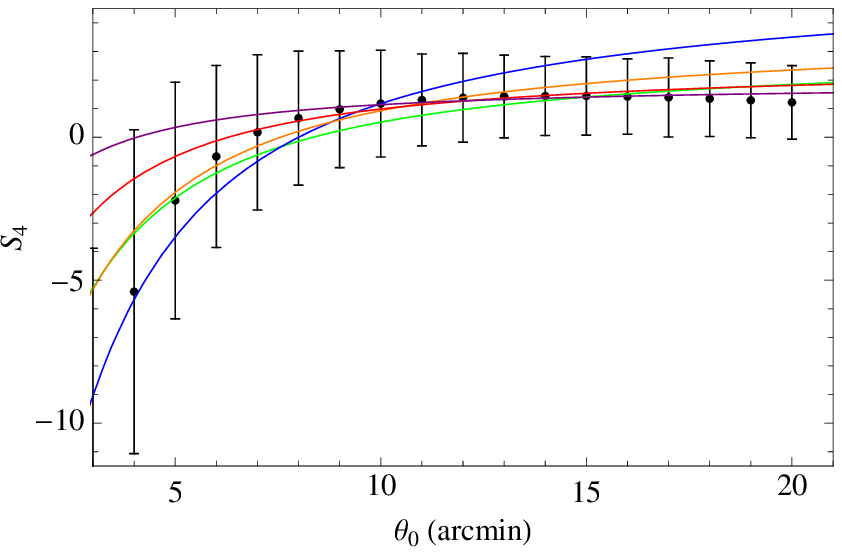} 
\caption{Same as Fig. \ref{fig: fitmomnonoise} but using denoising for the convergence (top panels) or the shear (bottom panels). Note that $\langle \kappa^4 \rangle$ becomes negative for small $\theta_0$ so that the corresponding values can not be shown in the logarithmic scale used for the moments plots.}
\label{fig: fitmomdenoise}
\end{figure*}

\begin{table*}
\begin{center}
\caption{Same as Table \ref{tab: mcvalsnonoise} but using denoising.} 
\resizebox{17cm}{!}{
\begin{tabular}{cccccccccccccccc}
\hline \hline
\multicolumn{16}{c}{No mass reconstruction - Yes noise - Yes denoising} \\
\hline
$\theta_0$ range & \multicolumn{3}{c}{(2 - 20)} & \multicolumn{3}{c}{(2 - 12)} & \multicolumn{3}{c}{(4 - 14)} & \multicolumn{3}{c}{(6 - 16)} & \multicolumn{3}{c}{8 - 18} \\
\hline
Id & $m$ & $c$ & $\rho_{rms}$ & $m$ & $c$ & $\rho_{rms}$  & $m$ & $c$ & $\rho_{rms}$  & $m$ & $c$  & $\rho_{rms}$  & $m$ & $c$ & $\rho_{rms}$  \\
\hline
$\mu_2$ & $-0.50 \pm 0.02$ & $-0.91 \pm 0.15$ & $3.94$ & $-0.53 \pm 0.02$ & $-0.48 \pm 0.27$ & $2.84$ & $-0.49 \pm  0.01$ & $-0.96 \pm 0.14$ & $1.38$ & $-0.46 \pm 0.01$ & $-1.18 \pm 0.07$ & $0.69$ & $-0.44 \pm 0.01$ & $-1.28 \pm 0.03$ & $0.29$ \\
$\mu_3$ & $-0.633 \pm 0.01$ & $-0.72 \pm 0.06$ & $5.65$ & $-0.63 \pm 0.01$ & $-0.85 \pm 0.17$ & $2.73$ & $-0.61 \pm 0.01$ & $-0.96 \pm 0.05$ & $1.29$ & $-0.62 \pm 0.01$ & $-0.85 \pm 0.07$ & $2.29$ & $-0.64 \pm 0.01$ & $-0.73 \pm 0.07$ & $2.97$ \\
$\mu_4$ & $-0.84 \pm 0.03$ & $0.29 \pm 0.27$ & $23.9$ & $-0.87 \pm 0.03$ & $2.01 \pm 1.01$ & $17.5$ & $-0.82 \pm 0.02$ & $0.64 \pm 0.43$ & $9.36$ & $-0.78 \pm 0.02$ & $0.08 \pm 0.19$ & $5.26$ & $-0.76 \pm 0.01$ & $-0.15 \pm 0.08$ & $2.75$ \\
$S_3$ & $0.56 \pm 0.04$ & $-0.28 \pm 0.02$ & $3.07$ & $0.52 \pm 0.06$ & $-0.26 \pm 0.03$ & $2.49$ & $0.66 \pm 0.03$ & $-0.32 \pm 0.01$ & $0.95$ & $0.72 \pm 0.01$ & $-0.34 \pm 0.01$ & $0.28$ & $0.72 \pm 0.03$ & $-0.34 \pm 0.01$ & $0.77$ \\
$S_4$ & $-0.73 \pm 0.02$ & $2.32 \pm 0.07$ & $2.44$ & $-0.76 \pm 0.03$ & $2.45 \pm 0.10$ & $2.01$ & $-0.71 \pm 0.02$ & $2.29 \pm 0.05$ & $0.70$ & $-0.69 \pm 0.01$ & $2.21 \pm 0.03$ & $0.39$ & $-0.66 \pm 0.01$ & $2.16 \pm 0.02$ & $0.22$ \\
\hline \hline
\multicolumn{16}{c}{Yes mass reconstruction - Yes noise - Yes denoising} \\
\hline
$\theta_0$ range & \multicolumn{3}{c}{(2 - 20)} & \multicolumn{3}{c}{(2 - 12)} & \multicolumn{3}{c}{(4 - 14)} & \multicolumn{3}{c}{(6 - 16)} & \multicolumn{3}{c}{8 - 18} \\
\hline
Id & $m$ & $c$ & $\rho_{rms}$ & $m$ & $c$ & $\rho_{rms}$  & $m$ & $c$ & $\rho_{rms}$  & $m$ & $c$  & $\rho_{rms}$  & $m$ & $c$ & $\rho_{rms}$  \\
\hline
$\mu_2$ & $-0.50 \pm 0.01$ & $-1.45 \pm 0.05$ & $1.48$ & $-0.50 \pm 0.01$ & $-1.50 \pm 0.09$ & $0.93$ & $-0.49 \pm 0.01$ & $-1.58 \pm 0.02$ & $0.31$ & $-0.49 \pm 0.01$ & $-1.53 \pm 0.04$ & $0.59$ & $-0.51 \pm 0.01$ & $-1.44 \pm 0.05$ & $0.76$ \\
$\mu_3$ & $-0.55 \pm 0.02$ & $-0.94 \pm 0.08$ & $6.76$ & $-0.53 \pm 0.02$ & $-1.18 \pm 0.27$ & $3.70$ & $-0.52 \pm 0.01$ & $-1.29 \pm 0.09$ & $1.88$ & $-0.54 \pm 0.01$ & $-1.11 \pm 0.09$ & $2.63$ & $-0.57 \pm 0.02$ & $-0.93 \pm 0.09$ & $3.19$ \\
$\mu_4$ & $-0.93 \pm 0.10$ & $0.37 \pm 0.67$ & $178$ & $-1.06 \pm 0.17$ & $3.96 \pm 3.92$ & $96.5$ & $-0.91 \pm 0.06$ & $0.91 \pm 0.86$ & $297$ & $-0.84 \pm 0.02$ & $0.00 \pm 0.26$ & $13.4$ & $-0.81 \pm 0.01$ & $-0.25 \pm 0.07$ & $4.08$ \\
$S_3$ & $0.83 \pm 0.17$ & $-0.29 \pm 0.07$ & $12.1$ & $0.59 \pm 0.21$ & $-0.17 \pm 0.10$ & $5.56$ & $1.01 \pm 0.12$ & $-0.35 \pm 0.05$ & $2.96$ & $1.28 \pm 0.13$ & $-0.45 \pm 0.05$ & $2.47$ & $1.51 \pm 0.10$ & $-0.53 \pm 0.04$ & $1.50$ \\
$S_4$ & $-2.11 \pm 0.56$ & $3.56 \pm 1.28$ & $110$ & $-2.96 \pm 0.95$ & $6.52 \pm 2.96$ & $175$ & $-2.19 \pm 0.36$ & $4.20 \pm 0.99$ & $79.6$ & $-1.69 \pm 0.21$ & $2.89 \pm 0.51$ & $37.1$ & $-1.33 \pm 0.18$ & $2.05 \pm 0.39$ & $14.1$ \\
\hline \hline
\end{tabular}}
\label{tab: mcvalsdenoise}
\end{center}
\end{table*}

\begin{displaymath}
\overline{\mu_4(\theta_0, {\rm noise}) /\mu_4(\theta_0, {\rm no \ noise})} = 0.194 \pm 0.003 \ .
\end{displaymath}

\begin{displaymath}
\overline{S_4(\theta_0, {\rm noise}) /S_4(\theta_0, {\rm no \ noise})} = 0.88 \pm 0.02  \ .
\end{displaymath}
The Gaussianization by the noise is again evident noting that, averaging over $\theta_0$, we get $\bar{S}_4 = 3.25$ to be compared to $\bar{S}_4 = 3.88$ for the noiseless case. Remembering that, for a Gaussian distribution it is $S_4 = 3$, it is therefore clear that adding a Gaussian noise pushes the convergence distribution towards a normal one thus lowering the kurtosis. Being $S_4 = \mu_4/\mu_2^2$ and having $\mu_2$ been decreased because of noise, $\mu_4$ has to decrease too in order to lead the ratio closer to the Gaussian value which is what we indeed find.

While the above results make it easier to grasp what the impact of noise is, in practical applications, one relies on the KS93 method to reconstruct the convergence map from noisy shear data. We therefore add random value of the intrinsic ellipticity on both the shear components in the MICECATv2 catalog and use these noisy data as input to the KS93 code to get the $\kappa$ maps we use to estimate moments, skewness and kurtosis. The results are shown in the bottom panels of Fig. \ref{fig: fitmomyesnoise} with the calibration parameters given in the lower half of Table \ref{tab: mcvalsyesnoise}.

Different from the noiseless case, we now find that the value of the $(m, c)$ parameters significantly depend on the fitting range. This can be seen looking at how the best fit curves are close to each other in the noiseless case, while they can be easily discriminated now. Discussing moments order by order, we start noting that the noise again decreases the $\mu_2(\theta_0)$ amplitude, but not the shape. We indeed get

\begin{displaymath}
\overline{\mu_2(\theta_0, {\rm noise}) /\mu_2(\theta_0, {\rm no \ noise})} = 0.68 \pm 0.07
\end{displaymath}
so that the effect is less dramatic. Again, it is not clear why the map reconstruction procedure alters the moments. It is possible that the KS93 method implicitly regularizes the convergence map thus smoothing out the high $\kappa$ tail and making the distribution more similar to a Gaussian one. 

However, the qualitative interpretation of noise as a Gaussianization agent does not refer to how the distribution have been obtained, but only to their final shape. We therefore expect similar results for the ratio of moments from noiseless and noisy reconstructed map. We indeed get 

\begin{displaymath}
\overline{\mu_3(\theta_0, {\rm noise}) /\mu_3(\theta_0, {\rm no \ noise})} = 0.28 \pm 0.02 \ , 
\end{displaymath}

\begin{displaymath}
\overline{\mu_4(\theta_0, {\rm noise}) /\mu_4(\theta_0, {\rm no \ noise})} = 0.31 \pm 0.02 \ , 
\end{displaymath}
which are the same order as the values obtained before. 

More involved is the situation for the skewness and kurtosis parameters. On one hand, we get

\begin{displaymath}
\overline{S_3(\theta_0, {\rm noise}) /S_3(\theta_0, {\rm no \ noise})} = 0.44 \pm 0.04 \ , 
\end{displaymath}

\begin{displaymath}
\overline{S_4(\theta_0, {\rm noise}) /S_4(\theta_0, {\rm no \ noise})} = 0.79 \pm 0.06 \ , 
\end{displaymath}
whic are smaller than before. However, the $\theta_0$ averaged values of both $S_3$ and $S_4$ again move towards their Gaussian counterparts with $(\bar{S}_3, \bar{S}_4) = (0.22, 3.24)$ instead of $(\bar{S}_3, \bar{S}_4) = (0.68, 4.84)$ for the noiseless case. 

Actually, the impact of noise on $(S_3, S_4)$ strongly depends on the smoothing length. Indeed, for $\theta_0 < 5 \ {\rm arcmin}$, the shape of the $S_3(\theta_0)$ and $S_4(\theta_0)$ curves is changed, not only the amplitude. As a result, both curves looses the monotonic behaviour with $\theta_0$ in favour of an increase with the smoothing length for $\theta_0 < 5 \ {\rm arcmin}$. As a consequence, fitting Eq.(\ref{eq: obsvsth}) to the full range gives bad results with large $\rho_{rms}$ values. This is a consequence of the theoretical model not including any contribution from the noise so that, as far as its impact is important, a linear matching is no more possible. As $\theta_0$ gets larger, the noise is more and more smoothed out so that the recovered $(S_3, S_4)$ becomes more and more similar to the theorerical values making the linear relation a good approximation. We therefore recommend to use only scales larger than 5 arcmin when comparing theory with observations ba
 sed on noisy data.

\section{Denoising}

As discussed above, noise (both on covergence and shear maps) works as a Gaussianization tool pushing moments towards the values expected for a normal distribution. Moreover, when the mass reconstruction procedure is applied and the smoothing length is smaller than 5 arcmin, it is not only the amplitude, but also the shape of the $S_3(\theta_0)$ and $S_4(\theta_0)$ changed. \cite{Ludo2013} have suggested a denoising procedure to remove the correlated noise contribution from the moments measured on the observed noisy convergence maps. In particular, the corrected moments read

\begin{equation}
\mu_{2}^{corr}(\theta_0) = \mu_{2}^{obs}(\theta_0) - \langle \mu_{2}^{rnd}(\theta_0) \rangle \ , 
\label{eq: mu2denoise}
\end{equation}

\begin{equation}
\mu_{3}^{corr}(\theta_0) = \mu_{3}^{obs}(\theta_0) \ ,
\label{eq: mu3denoise}
\end{equation}

\begin{equation}
\mu_{4}^{corr}(\theta_0) = \mu_{4}^{obs}(\theta_0) - 6 \langle \mu_{2}^{obs}(\theta_0) \mu_{2}^{rnd}(\theta_0) \rangle
- \langle \mu_{4}^{rnd}(\theta_0) \rangle \ ,
\label{eq: mu4denoise}
\end{equation}
where quantities with the label $obs$ are measured over the observed map, while the label $rnd$ refers to noise only maps and are estimated by averaging over a large number of realizations. Since there are no explicit formulae for the skewness and kurtosis, we compute their denoised equivalents by simply plugging denoised quantities in their definitions.

We first build 100 noise only maps for both convergence and shear denoising. The noise convergence maps have been realized randomizing convergence values, but keeping fixed galaxies positions.When using map reconstruction, we instead keep fixed both position and shape of galaxies,  but randomize their orientation. For a given filter and smoothing length, we then use these maps to estimate the terms entering Eqs.(\ref{eq: mu2denoise}) and (\ref{eq: mu4denoise}). Note that no correction is needed for the third order moment, but denoising still affects the skewness because of the change in second order moment.

Let us first consider the case with no mass reconstruction and noise added directly to the convergence map (see top panels in Fig. \ref{fig: fitmomdenoise} and upper half of Table \ref{tab: mcvalsdenoise}). Somewhat surprisingly, we find that denoising has almost no effect at all. Indeed, both the moments and the skewness and kurtosis parameters are so close to their noisy counterparts that no noticeable impact on the relation with noiseless quantities and the calibration coefficients $(m, c)$ can be noticed. This is likely a consequence of having averaged the noise moments over 100 realizations so that their final contribution is lowered. Indeed we find that the corrective terms on the right hand side of Eqs.(\ref{eq: mu2denoise}) and (\ref{eq: mu4denoise}) are of order $1\%$ ($2\%$) for $\mu_2$ $(\mu_4)$ so that denoising is actually not working at all. However, we again stress that working on convergence map is an idealized condition which never apply in realistic situatio
 n so that we do not worry about such a failure. 

On the contrary, when map reconstruction with the KS93 method is performed, denoising starts having a serious impact. Indeed, now the corrective terms are of order $33\%$ ($60\%$) for $\mu_2$ ($\mu_4$) so that we now get

\begin{displaymath}
\overline{\mu_2(\theta_0, {\rm denoise}) /\mu_3(\theta_0, {\rm no \ noise})} = 0.88 \pm 0.01 \ , 
\end{displaymath}

\begin{displaymath}
\overline{\mu_4(\theta_0, {\rm denoise}) /\mu_4(\theta_0, {\rm no \ noise})} = 0.17 \pm 0.02 \ . 
\end{displaymath}
These numbers nevertheless point at a failure of the denoising procedure since one should actually find values of order unity, i.e. the denoised moments match the noiseless ones. On the contrary, such low numbers are an evidence that denoising is removing both the noise and (part of) the signal. A qualitative explanation for this failure can be traced back considering how Eqs.(\ref{eq: mu2denoise})\,-\,(\ref{eq: mu4denoise}) have been obtained. It is easy to show that they are indeed exact if both the noise and the input $\kappa$ distributions are Gaussian, but this is actually not the case because of the positively skewed nature of the convergence distribution. As such, making the normal distribution assumption indeed removes the high $\kappa$ tails thus cancelling its asymmetry. Indeed, we also find

\begin{displaymath}
\overline{S_3(\theta_0, {\rm denoise}) /S_3(\theta_0, {\rm no \ noise})} = 0.65 \pm 0.04 \ , 
\end{displaymath}
which is consistent with denoising forcing the distribution to be more symmetric than it actually is. As a further consequence, the kurtosis is so strongly reduced that it becomes negative for smoothing length smaller than $10 \ {\rm arcmin}$ so that fitting a linear relation between reconstruced and theoretical moments provide bad results. 

As a possible way out, one could try to empirically modify Eqs.(\ref{eq: mu2denoise})\,-\,(\ref{eq: mu4denoise}) adding some multiplicative factors to the correction terms and tailoring them so that the noiseless moments are exactly recovered. However, this would ask for a large number of both noise and signal simulations to be sure that these corrective factors do not depend on the noise properties and the cosmological parameters, but are rather universal. This is outside our aims here so that we conservatively argue against using denoising in moments estimate.

\section{Discussion}

Having presented the results on estimating moments from simulated maps with and without noise and/or map reconstruction, it is time to draw some general lessons and consider what is still missing from this first analysis.

\subsection{Can we use moments?}

The main aim of this work was to investigate whether and to which extent high order moments of lensing convergence can be used to infer constraints on cosmological parameters. While the advantage of using moments will be quantitified in Paper II though a Fisher matrix forecast analysis, the preliminary question to solve was to look for a reliable way to connect theory and observations. Indeed, a whatever observable can be a valid help for cosmology if the relation between what is measured and what is predicted is well defined. In the case of moments, a one\,-\,to\,-\,one matching is not possible both because of observational problems (i.e., what we measure is not a perfect tracer of the quantity of interest) and theoretical issues (mainly because of uncertainties in the derivation of convergence bispectrum and nonlinearities, but see Paper II for details). However, moments can still be used if the proposed linear relation (\ref{eq: obsvsth}) is shown to be valid under realist
 ic situations. 

To answer this question, one must look at the lower half part of Table \ref{tab: mcvalsyesnoise} which refers to what is typically done with observations, i.e. a convergence map is reconstructed applying the KS93 method on noisy data. Considering that moments are measured with an error no smaller than $30\%$ (depending on the order and the smoothing length used), we can conservatively deem as acceptable the linear mapping from theoretical to observed quantities if the rms percentage of the residuals is smaller than $\sim 10\%$. Looking at the $\rho_{rms}$ values, we can conclude that the scaling of moments with the smoothing length $\theta_0$ can indeed be used as a cosmological tool provided the range $2 < \theta_0/{\rm arcmin} < 4$  is cut out. Whether it is better to use the set $(\mu_2, \mu_3, \mu_4)$ or $(\mu_2, S_3, S_4)$ is a problem we will address in Paper II. 

Although we have argued against denoising becasue of the problem with the fourth order moment, a mixed strategy could be possible. Indeed, denoising improves the accuracy of the linear fit for both $\mu_2$ and $S_3$ so that also the range $(2, 4) 	\ {\rm arcmin}$ becomes usable. One could therefore take as data the denoised $\mu_2(\theta_0)$ and $S_3(\theta_0)$, while retaining the noisy $S_4(\theta_0)$. We will therefore investigate in Paper II this (somewhat contradictory) approach too.

\subsection{What is still missing?}

The reason why we have added noise to the shear components before using the KS93 method to reconstruct the convergence map was to simulate realistic data. Although this is a first step in the right direction, further ones are needed. 

First, we have assumed that the full survey area is observable. Actually, in order to remove stars and other artifacts, part of each filed will be masked. Masking has a double impact on moments estimate. On one hand, it reduces the effective survey area thus making harder to fulfill the requirement in Sect. 2.2. Second, it cuts pixels in the neighbourhood of masked zones thus cancelling the signal. Even if a detailed simulation taking care of both the stars distribution over the sky and the instrumental setup is needed, we nevertheless speculate that masking should not prevent or bias the estimate of moments. Indeed, moments are global quantities related to the properties of the convergence field over the map rather than to any local features. As far as masking does not correlate with $\kappa$ (e.g., if high $\kappa$ pixels are preferentially masked instead of low $\kappa$ ones), we do not expect any change in the estimated moments. However, reducing the area reduction could 
 increase the error on the measured moments, an effect which is hard to quantify. 

A second problem concerns the source number density $n_g$ and the source redshift distribution. We have implicitly assumed that they are the same all over the survey area so that the moments from different fields are different representations of the same underlying quantities which can then be averaged to wash out peculiarities of each single map (e.g., voids or clustered structures). Because of the different sky background, the magnitude limit will be different from one field to another thus making the redshift distribution a function of the position of the map on the sky. However, as long as the redshift distribution does not abruptly change from one field to another, we do not expect any siginificant systematics to affect the esimated moments so that any small variation can be taken into account phenomenologically through the parameterization given by Eq.(\ref{eq: obsvsth}). That this is the case, it is already confirmed by the data we have used. Indeed, the redshift distr
 ibution for each of the 140 fields we have used scatters around the average one of the full MICECATv2 catalog and yet we get a very good fit. 

\subsection{What about intrinsic alignment?}

It is worth taking a step back to the very beginning to understand how intrinsic alignment (IA) may possibly affect the estimate of moments. The starting point is the MICECATv2 catalog with the positions of galaxies which we then project on a gridded plane. Since we work in the weak lensing regime, the observed ellipticity ${\bf e}$ of a galaxy is the sum of the intrinsinc one ${\bf e}_{int}$ and the shear $\gamma$ so that, under the assumption of random orientation, averaging over the large number of galaxies in a given pixel and assuming the shear is constant over the pixel area (so that it is the same for all galaxies), we can attach to each pixel an estimate of the shear by averaging over the galaxies being 

\begin{displaymath}
\langle {\bf e} \rangle = \langle {\bf e}_{int} \rangle + \gamma = \gamma \ .
\end{displaymath}
If IA is present, the above relation is no more valid so that the measured shear is a biased estimate of the actual one, i.e. $\gamma_{obs} = \gamma + \gamma_{IA}$. Since Fourier transform is a linear operation and the KS93 method is indeed based on Fourier transforming the shear field, one should get a biased reconstruction of the convergence field with $\kappa_{rec} = \kappa + \kappa_{IA}$ and $\kappa_{IA}$ the term corresponding to $\gamma_{IA}$. From the point of view of moments, this would not be dramatic only amounting to a shift of the centre of the convergence distribution, but not its shape. As such, one could naively expect that the skewness and kurtosis parameters are unaffected, while the second order moment is changed because of the dispersion in the $\gamma_{IA}$ values. Actually, things are much more complicated. Fisrt, IA is not constant over the field so that the $\kappa_{IA}$ term change from one pixel to another introducing features in the reconstruction wh
 ich changes the final shape of the $\kappa$ distribution and hence its moments. Second, IA has not a white noise\,-\,like power spectrum so that it biases the measured convergence power spectrum thus again changing the statistical properties of the $\kappa$ distribution. Quantifying this effect requires adding to each galaxy an intrinsic ellipitcity in such a way that the IA power spectrum is reproduced. This is outside our aim here, but it is an issue which should carefully be addressed to in a forthcoming work.

As a final remark, we note that an IA contamination could possibly be found considering how the smoothing and KS93 method works on noisy data. In presence of noise, the shear in each pixel reads

\begin{displaymath}
\gamma_{pix} = \gamma + \gamma_N
\end{displaymath}
with $\gamma_N$ the noise contribution. In the usual approach, one first smooth the shear field and then perform map reconstruction through the KS93 method. Smoothing cancels the $\gamma_N$ term so that the KS93 technique gives $\kappa_{pix}$ from $\gamma_{pix}$. It is immediate to understand that the same result is obtained if one inverts the two operations. Indeed, should we first apply the KS93 method, we would obtain

\begin{displaymath}
\kappa_{pix} = \kappa + \kappa_N 
\end{displaymath}
with $\kappa_N$ the convergence term corresponding to $\gamma_N$. Since the noise is uncorrelated with the signal and is purely random, smoothing the convergence field thus obtained cancels the $\kappa_N$ term thus leaving the same convergence field as before. Although this simplified mathematical demonstration is convincing enough, we have nevertheless verified that smoothing and reconstrucion are commutative properties.

However, should IA be present, the input shear reads

\begin{displaymath}
\gamma_{pix} = \gamma  + \gamma_{IA} + \gamma_N
\end{displaymath}
which, after smoothing, reduces to 

\begin{displaymath}
\gamma_{sm} = \gamma  + f_{sm} \gamma_{IA}
\end{displaymath}
with $f_{sm}$ taking care of the partial cancellation of the IA term. The KS93 method now gives

\begin{displaymath}
\kappa_{SR} = \kappa + f_{sm} \kappa_{IA} + \kappa_{\gamma I}
\end{displaymath}
with $\kappa_{\gamma I}$ the term arising from the correlation between shear and IA. Suppose now we first make KS93 reconstruction and then perform smoothing. The final convergence field will be

\begin{displaymath}
\kappa_{RS} = \kappa + \kappa_{IA} + f_{sm}^{\prime} \kappa_{\gamma I} \ne \kappa_{SR} \ .
\end{displaymath}
Even if this is a very qualitative derivation, we can nevertheless argue that the presence of IA makes the smoothing and reconstruction no more commutative. Quantifying the difference between $\kappa_{SR}$ and $\kappa_{RS}$ asks for simulations including the IA term which is not possible with present data.

\section{Conclusions}

Future Stage IV surveys, such as Euclid and LSST, will allow lensing to ascend the ladder towards a full maturity as a cosmological tool. Moreover, the unprecedented image quality and the large area will make it worth asking whether it is possible to make the next step forward going beyond second order statistics. The promise to break degeneracy among cosmological parameters and hence increase the Figure of Merit makes it worth trying to override the observational and theoretical hurdles related to higher order statistics. 

To this end, rather than dealing with the two components shear field $\gamma$, it is more convenient to rely on a scalar field such as the lensing convergence $\kappa$ which can be reconstructed from $\gamma$ trhough the Fourier transform based KS93 method. Working with $\kappa$ greatly simplifies both the theoretical derivation and the observational measurement. We have therefore start a program under which conditions and to which extent up to fourth order convergence moments can boost the Figure of Merit when combined with cosmic shear tomography. The present paper first addresses the problem of whether moments measured from convergence maps can actually be used at all. In order this to be the case, one should demonstrate that there is a one\,-\,to\,-\,one relation between observed and theoretical moments. Through the use of shear and convergence maps extracted from the MICE Grand Challenge simulation, we have shown that a linear relation is sufficient to match observations
  and theory thus making high order convergence moments a reliable tool. 

Through a step\,-\,by\,-\,step analysis, we have shown that such a linear approximation holds not only when the actual convergence map is used (which is actually not possible since one has no direct access to it), but also when the $\kappa$ field is reconstructed from noisy $\gamma$ data. Our extensive analysis has shown that the skewness $S_3$ and the kurtosis $S_4$ can more reliably be matched with theory since they can be calibrated with a smaller scatter. Moreover, we have also argued against a blind use of denoising since this procedure does not recover the correct dependence of $(S_3, S_4)$ on the aperture of the filter used for smoothing the map.

Although encouraging, these results should still be considered preliminary. First, we note that the MICE lightcone data we have used mainly probe the redshift range $0.1 \le z \le 1.4$, while future surveys will push the upper limit to $z \sim 2.5$ thus making it important to extend our analysis to such large $z$. This can only be done relying on a different simulated dataset such as the one which could be obtained using the recently released Multidark weak lensing lightcone \citep{Gio15}. This will offer the possibility to not only extend the redshift range, but also adjust the source redshift distribution thus allowing to explore the impact of this quantity on the validity of our linear calibration relation. Moreover, being Multidark based on a different input cosmological model, we could also check whether the results in the present paper still hold when the background cosmology is changed. We stress that, in order for a such a test to be passed with green lights, it is no
 t important that the calibration coefficients $(m, c )$ are the same since they will likely depend on both cosmology and the redshift range probed, but only that the relation between theoretical and observed moments can be approximated as a linear one with a small scatter. Although such a confirmation test is needed,  we are nevertheless confident that this is indeed the case since there is nothing in our procedure related to which cosmological model is the input for the simulation.

While we have added noise to the shear maps in order to mimic real data, something is still missing in order to make them as close as possible to what is actually measured from images. Indeed, we have here implicitly assumed that the shape measurement error is negligible and no systematics are present. Indeed, shape measurement codes are improving more and more their efficiency, yet they are still not perfect. As a consquence, the measured shape is a noisy estimator of the true galaxy ellipticity and this error should be propagated somewhat first in the reconstruction procedure and then to the measured moments. How to do this is unfortunately not well understood. First, we note that present day shear catalogs typically provide estimate of ellipticity and a weight taking care of both the statistical uncertainty of the estimator and the intrinsic ellipticity variance. Separating the two components is not so immediate since the combination of the two terms is tailored on the spe
 cifics of the survey and the shape measurement method. However, no matter which code is used, the shape measurement error is typically smaller thant the intrinsic ellipticity variance provided the galaxy signal\,-\,to\,-\,noise ratio $S/N$ is large enough. This will likely be the case for the majority of the galaxies which will make the sample of future lensing surveys so that we are confident that the shape measurement noise will be negligible with respect to the intrinsic ellipticity noise we have already included in the present work. 

Systematic errors in the galaxy ellipticity determination can affect higher order moments too. Assuming they are sufficiently small, one can write $\gamma_{obs} = (1 + m) \gamma + c$ with $(m, c)$ the multiplicative and additive bias. Should they be uncorrelated with the shear, it is easy to show that, because of the linearity of Fourier transform, the reconstructed $\kappa$ field will be a biased representation of the actual so that $\kappa_{obs} = (1 + m_{\kappa}) \kappa + c_{\kappa}$ with $(m_{\kappa}, c_{\kappa})$ related to $(m, c)$. Should this be the case, the convergence probability distribution will be miscentred, but its shape is unchanged so that $(\mu_2, S_3, S_4)$ should not be biased. Actually, $(m, c)$ can be scale and redshift dependent so that their impact on the moments is far from trivial. In order to quantify it, one should attach $(m, c)$ values to each pixel in the shear map and perform the full moments estimate starting from these biased $\gamma$ fields
 . Unfortunately, how to choose $(m, c)$ is unclear. While some approximated formulae exist giving $m$ as a function of the galaxy $S/N$ and size, they are again strongly dependent on the shape measurement code and the simulations used to calibrate the method. Extrapolating to a different survey can be risky so that the impact of systematics will likely require an ad hoc end\,-\,to\,-\,end analysis.   

Even if far from conclusive, the results presented in this paper convincingly show that going beyond the second order statistics is indeed possible. In the companion Paper II, we will show that this is also a rewarding step up. 

 \section*{Acknowledgments}

The authors acknowledge E. Jullo, T. Kitching, and S. Pires for comments on a preliminary version of the manuscript. VFC also thanks F. Bernardeau for discussion on higher order perturbation theory. We acknowledge the use of data from the MICE simulations, publicly available at {\tt http://www.ice.cat/mice}; in particular, we thank the MICE team for allowing us to use the MICECAT v2.0 catalog prior of publication. VFC and XE are funded by Italian Space Agency (ASI) through contract Euclid\,-\,IC (I/031/10/0) and acknowledge financial contribution from the agreement ASI/INAF/I/023/12/0.


\appendix

\section{Top hat filter}

Table\,\ref{tab: mcvalstophat} summarizes the values of the $(m, c)$ calibration coefficients and the rms of the percentage residuals for the different cases discussed in the text adopting a top hat rather than a Gaussian filter. Although the exact values are different, we can nevertheless note that the linear approximation still works quite well so that all what we have said before still qualitatively holds. Choosing which filter should be adopted when measuring moments can therefore not be motivated by a measurement point of view. We will refer the reader to Paper II where we will investigate whether the constraints on cosmological parameters from a combined use of moments and shear tomography depend on the filter adopted.

\begin{table*}
	\begin{center}
		\caption{Calibration parameters, $(m, c)$, and rms percentage residuals, $\rho_{rms} = rms[100 \times (1 - \mu_n(fit)/\mu_n(obs))]$, for different moments. Note that $c$ is in units of $(10^{-6}, 10^{-9}, 10^{-11})$ for $(\mu_2, \mu_3, \mu_4)$, respectively. A Top Hat filter is used to smooth the map.} 
		\resizebox{17cm}{!}{
			\begin{tabular}{cccccccccccccccc}
				\hline \hline
				\multicolumn{16}{c}{No mass reconstruction - No noise - No denoising} \\
				\hline
				$\theta_0$ range & \multicolumn{3}{c}{(2 - 20)} & \multicolumn{3}{c}{(2 - 12)} & \multicolumn{3}{c}{(4 - 14)} & \multicolumn{3}{c}{(6 - 16)} & \multicolumn{3}{c}{8 - 18} \\
				\hline
				Id & $m$ & $c$ & $\rho_{rms}$ & $m$ & $c$ & $\rho_{rms}$  & $m$ & $c$ & $\rho_{rms}$  & $m$ & $c$  & $\rho_{rms}$  & $m$ & $c$ & $\rho_{rms}$  \\
				\hline
				$\mu_2$ & $-0.12 \pm 0.04$ & $0.33 \pm 0.56$ & $3.84$ & 
				$-0.18 \pm 0.04$ & $1.78 \pm 0.76$ & $2.17$  & $-0.10 \pm 0.03$ & $0.36 \pm 0.49$ & $1.30$ & $-0.03 \pm 0.03$ & $-0.53 \pm 0.36$ & $0.91$ & $0.02 \pm 0.02$ & $-1.18 \pm 0.27$ & $0.66$ \\
				$\mu_3$ & $0.42 \pm 0.07$ & $-3.14 \pm 1.39$ & $6.35$ & $0.34 \pm 0.09$ & $1.95 \pm 4.07$ & $5.57$ & $0.51 \pm 0.05$ & $-3.52 \pm 1.47$ & $2.37$ & $0.59 \pm 0.02$ & $-5.31 \pm 0.50$ & $0.92$ & $0.61 \pm 0.01$  & $-5.73 \pm 0.13$ & $0.32$ \\
				$\mu_4$ & $-0.38 \pm 0.09$ & $14.5 \pm 5.43$ & $17.5$ & $-0.47 \pm 0.08$ & $41.7 \pm 14.9$ & $12.1$ & $-0.32 \pm 0.07$ & $22.2 \pm 7.10$ & $7.27$ & $-0.21 \pm 0.06$ & $12.7 \pm 3.92$ & $4.82$ & $-0.11 \pm 0.05$ & $7.18 \pm 2.22$ & $3.30$ \\
				$S_3$ & $1.01 \pm 0.09$ & $-0.35 \pm 0.05$ & $4.29$ & $0.87 \pm 0.11$ & $-0.25 \pm 0.07$ & $2.31$ & $1.10 \pm 0.08$ & $-0.39 \pm 0.04$ & $1.32$ & $1.25 \pm 0.06$ & $-0.46 \pm 0.03$ & $0.87$ & $1.37 \pm 0.04$ & $-0.51 \pm 0.02$ & $0.51$ \\
				$S_4$ & $-0.56 \pm 0.01$ & $2.24 \pm 0.05$ & $1.34$ & $-0.58 \pm 0.01$ & $2.40 \pm 0.10$ & $1.04$ & $-0.56 \pm 0.01$ & $2.24 \pm 0.05$ & $0.54$ & $-0.54 \pm 0.01$ & $2.16 \pm 0.03$ & $0.28$ & $-0.53 \pm 0.01$ & $2.12 \pm 0.01$ & $0.11$ \\
				\hline \hline
				\multicolumn{16}{c}{Yes mass reconstruction - No noise - No denoising} \\
				\hline
				$\theta_0$ range & \multicolumn{3}{c}{(2 - 20)} & \multicolumn{3}{c}{(2 - 12)} & \multicolumn{3}{c}{(4 - 14)} & \multicolumn{3}{c}{(6 - 16)} & \multicolumn{3}{c}{8 - 18} \\
				\hline
				Id & $m$ & $c$ & $\rho_{rms}$ & $m$ & $c$ & $\rho_{rms}$  & $m$ & $c$ & $\rho_{rms}$  & $m$ & $c$  & $\rho_{rms}$  & $m$ & $c$ & $\rho_{rms}$  \\
				\hline
				$\mu_2$ & $-0.00 \pm 0.02$ & $-2.66 \pm 0.26$ & $1.85$ & $-0.04 \pm 0.03$ & $-1.92 \pm 0.50$ & $1.43$ & $-0.01 \pm 0.01$ & $-2.73 \pm 0.22$ & $0.60$ & $0.04 \pm 0.01$ & $-3.07\pm 0.12$ & $0.33$ & $0.05 \pm 0.01$ & $-3.25 \pm 0.06$ & $0.18$ \\
				$\mu_3$ & $1.12 \pm 0.05$ & $-9.87 \pm 0.71$ & $3.17$ & $1.14 \pm 0.07$ & $-10.4 \pm 2.84$ & $2.68$ & $1.22 \pm 0.02$ & $-12.5 \pm 0.44$ & $0.63$ & $1.19 \pm 0.04$ & $-11.7 \pm 0.72$ & $1.27$ & $1.12 \pm 0.05$ & $-10.2 \pm 0.81$ & $1.71$  \\
				$\mu_4$ & $0.00 \pm 0.09$ & $1.19 \pm 3.46$ & $11.9$ & $-0.10 \pm 0.10$ & $20.9 \pm 14.1$ & $10.1$ & $0.05 \pm 0.06$ & $4.71 \pm 5.23$ & $4.82$ & $0.14 \pm 0.04$ & $-1.51 \pm 2.02$ & $2.31$ & $0.18 \pm 0.02$ & $-3.68 \pm 0.64$ & $0.94$ \\
				$S_3$ & $1.88 \pm 0.12$ & $-0.52 \pm 0.07$ & $3.96$ & $1.71 \pm 0.16$ & $-0.40 \pm 0.10$ & $2.46$ & $2.07 \pm 0.09$ & $-0.60 \pm 0.05$ & $1.06$ & $2.25 \pm 0.06$ & $-0.68 \pm 0.03$ & $0.60$ & $2.35 \pm 0.04$ & $-0.73 \pm 0.02$ & $0.30$ \\
				$S_4$ & $-0.28 \pm 0.02$ & $2.05 \pm 0.10$ & $1.73$ & $-0.31 \pm 0.04$ & $2.24 \pm 0.24$ & $1.94$ & $-0.24 \pm 0.01$ & $1.89 \pm 0.06$ & $0.47$ & $-0.23 \pm 0.01$ & $1.84 \pm 0.02$ & $0.13$ & $-0.24 \pm 0.01$ & $1.88 \pm 0.04$ & $0.24$ \\
				\hline \hline
	
					\multicolumn{16}{c}{No mass reconstruction - Yes noise - No denoising} \\
					\hline
					$\theta_0$ range & \multicolumn{3}{c}{(2 - 20)} & \multicolumn{3}{c}{(2 - 12)} & \multicolumn{3}{c}{(4 - 14)} & \multicolumn{3}{c}{(6 - 16)} & \multicolumn{3}{c}{8 - 18} \\
					\hline
					Id & $m$ & $c$ & $\rho_{rms}$ & $m$ & $c$ & $\rho_{rms}$  & $m$ & $c$ & $\rho_{rms}$  & $m$ & $c$  & $\rho_{rms}$  & $m$ & $c$ & $\rho_{rms}$  \\
					\hline
					$\mu_2$ & $-0.53 \pm 0.01$ & $-0.44 \pm 0.15$ & $2.28$ & $-0.54 \pm 0.01$ & $0.01 \pm 0.15$ & $0.92$ & $-0.53 \pm 0.01$ & $-0.29 \pm 0.17$ & $0.95$ & $-0.50 \pm 0.01$ & $-0.62 \pm  0.14$ & $0.73$ & $-0.48 \pm 0.01$ & $-0.87 \pm 0.11$ & $0.55$ \\
					$\mu_3$ & $-0.65 \pm 0.02$ & $-0.61 \pm 0.30$ & $5.68$ & $-0.67 \pm 0.02$ & $0.45 \pm 0.89$ & $4.97$ & $-0.63 \pm 0.01$ & $-0.73 \pm 0.29$ & $1.97$ & $-0.61 \pm 0.01$ & $-1.08 \pm 0.09$ & $0.71$ & $-0.61 \pm 0.01$ & $-1.16 \pm 0.02$ & $0.17$ \\
					$\mu_4$ & $-0.89 \pm 0.02$ & $3.16 \pm 0.95$ & $17.3$ & $-0.90 \pm 0.01$ & $8.52 \pm 2.43$ & $9.97$ & $-0.88 \pm 0.01$ & $5.01 \pm 1.38$ & $7.49$ & $-0.86 \pm 0.01$ & $2.98 \pm 0.79$ & $5.35$ & $-0.83 \pm 0.01$ & $1.77 \pm 0.51$ & $3.99$ \\
					$S_3$ & $0.34 \pm 0.06$ & $-0.19 \pm 0.03$ & $4.19$ & $0.25 \pm 0.07$ & $-0.14 \pm 0.05$ & $2.29$ & $0.41 \pm 0.05$ & $-0.23 \pm 0.03$ & $1.23$ & $0.51 \pm 0.04$ & $-0.28 \pm 0.02$ & $0.71$ & $0.57 \pm 0.02$ & $-0.31 \pm 0.01$ & $0.42$ \\
					$S_4$ & $-0.77 \pm 0.01$ & $2.43 \pm 0.05$ & $1.54$ & $-0.78 \pm 0.01$ & $2.57 \pm 0.08$ & $1.07$ & $-0.76 \pm 0.01$ & $2.44 \pm 0.05$ & $0.57$ & $-0.75 \pm 0.01$ & $2.37 \pm 0.03$ & $0.31
					$ & $-0.74 \pm 0.01$ & $2.32 \pm 0.02$ & $0.22$ \\
					\hline \hline
					\multicolumn{16}{c}{Yes mass reconstruction - Yes noise - No denoising} \\
					\hline
					$\theta_0$ range & \multicolumn{3}{c}{(2 - 20)} & \multicolumn{3}{c}{(2 - 12)} & \multicolumn{3}{c}{(4 - 14)} & \multicolumn{3}{c}{(6 - 16)} & \multicolumn{3}{c}{8 - 18} \\
					\hline
					Id & $m$ & $c$ & $\rho_{rms}$ & $m$ & $c$ & $\rho_{rms}$  & $m$ & $c$ & $\rho_{rms}$  & $m$ & $c$  & $\rho_{rms}$  & $m$ & $c$ & $\rho_{rms}$  \\
					\hline
					$\mu_2$ & $0.97 \pm 0.38$ & $-14.4 \pm 4.59$ & $25.9$ & $1.58 \pm 0.62$ & $-26.7 \pm 11.0$ & $20.3$ & $0.89 \pm 0.25$ & $-15.4 \pm 3.77$ & $9.81$ & $0.47 \pm 0.13$ & $-9.75 \pm  1.62$ & $5.01$ & $0.26 \pm 0.08$ & $-7.20 \pm 0.91$ & $3.18$ \\
					$\mu_3$ & $-0.61 \pm 0.05$ & $-0.55 \pm 0.80$ & $19.2$ & $-0.66 \pm 0.08$ & $2.01 \pm 2.98$ & $19.7$ & $-0.55 \pm 0.02$ & $-1.09 \pm 0.50$ & $2.80$ & $-0.53 \pm 0.01$ & $-1.59 \pm 0.06$ & $0.40$ & $-0.53 \pm 0.01$ & $ -1.59 \pm  0.04$  & $0.37$ \\
					$\mu_4$ & $-0.32 \pm 0.15$ & $-8.39 \pm 5.57$ & $27.2$ & $-0.14 \pm 0.27$ & $-36.1 \pm 33.5$ & $27.2$ & $-0.33 \pm 0.09$ & $-14.3 \pm  7.20$ & $11.8$ & $-0.45 \pm 0.04$ & $-6.56 \pm 2.10$ & $5.46$ & $-0.50 \pm 0.02$ & $-3.93 \pm 0.84$ & $3.17$ \\
					$S_3$ & $-1.46 \pm 0.06$ & $0.45 \pm 0.05$ & $16.9$ & $-1.55 \pm 0.04$ & $0.53 \pm 0.03$ & $6.87$ & $-1.39 \pm 0.09$ & $0.43 \pm 0.05$ & $5.80$ & $-1.15 \pm 0.09$ & $0.30 \pm 0.04$ & $3.55$ & $-0.96 \pm 0.09$ & $0.22 \pm 0.04$ & $2.66$ \\
					$S_4$ & $-1.02 \pm 0.01$ & $3.45 \pm 0.09$ & $3.24$ & $-1.03 \pm 0.01$ & $3.61 \pm 0.07$ & $1.51$ & $-1.01 \pm 0.01$ & $3.44 \pm 0.11$ & $1.40$ & $-0.97 \pm 0.01$ & $3.22 \pm 0.08$ & $0.83$ & $-0.93 \pm 0.01$ & $3.07 \pm 0.06$ & $0.53$ \\
					\hline \hline
	
					\multicolumn{16}{c}{No mass reconstruction - Yes noise - Yes denoising} \\
					\hline
					$\theta_0$ range & \multicolumn{3}{c}{(2 - 20)} & \multicolumn{3}{c}{(2 - 12)} & \multicolumn{3}{c}{(4 - 14)} & \multicolumn{3}{c}{(6 - 16)} & \multicolumn{3}{c}{8 - 18} \\
					\hline
					Id & $m$ & $c$ & $\rho_{rms}$ & $m$ & $c$ & $\rho_{rms}$  & $m$ & $c$ & $\rho_{rms}$  & $m$ & $c$  & $\rho_{rms}$  & $m$ & $c$ & $\rho_{rms}$  \\
					\hline
					$\mu_2$ & $-0.57 \pm 0.02$ & $-0.01 \pm 0.28$ & $4.02$ & $-0.60 \pm 0.02$ & $0.75 \pm 0.41$ & $2.42$ & $-0.56 \pm  0.01$ & $0.01 \pm 0.25$ & $1.38$ & $-0.52 \pm 0.01$ & $-0.45 \pm 0.17$ & $0.91$ & $-0.50 \pm 0.01$ & $-0.75 \pm 0.13$ & $0.65$ \\
					$\mu_3$ & $-0.65 \pm 0.02$ & $-0.61 \pm 0.30$ & $5.68$ & $-0.67 \pm 0.02$ & $0.45 \pm 0.89$ & $4.97$ & $-0.63 \pm 0.01$ & $-0.73 \pm 0.29$ & $1.97$ & $-0.61 \pm 0.01$ & $-1.08 \pm 0.09$ & $0.71$ & $-0.61 \pm 0.01$ & $-1.17 \pm 0.02$ & $0.17$ \\
					$\mu_4$ & $-0.91 \pm 0.02$ & $3.78 \pm 1.31$ & $28.1$ & $-0.93 \pm 0.02$ & $10.6 \pm 4.17$ & $20.8$ & $-0.89 \pm 0.01$ & $5.41 \pm 1.56$ & $9.12$ & $-0.86 \pm 0.01$ & $3.14 \pm 0.84$ & $5.94$ & $-0.84 \pm 0.01$ & $-1.84 \pm 0.53$ & $4.29$ \\
					$S_3$ & $0.46 \pm 0.04$ & $-0.24 \pm 0.02$ & $2.73$ & $0.40 \pm 0.05$ & $-0.20 \pm 0.03$ & $1.48$ & $0.50 \pm 0.04$ & $-0.26 \pm 0.2$ & $0.91$ & $0.58 \pm 0.03$ & $-0.30 \pm 0.01$ & $0.51$ & $0.63 \pm 0.02$ & $-0.32 \pm 0.01$ & $0.29$ \\
					$S_4$ & $-0.86 \pm 0.05$ & $2.76 \pm 0.24$ & $13.1$ & $-0.92 \pm 0.08$ & $3.24 \pm 0.54$ & $13.0$ & $-0.80 \pm 0.02$ & $2.59 \pm 0.11$ & $1.41$ & $-0.77 \pm 0.01$ & $2.43 \pm 0.05$ & $0.49$ & $-0.75 \pm 0.01$ & $2.35 \pm 0.03$ & $0.29$ \\
					\hline \hline
					\multicolumn{16}{c}{Yes mass reconstruction - Yes noise - Yes denoising} \\
					\hline
					$\theta_0$ range & \multicolumn{3}{c}{(2 - 20)} & \multicolumn{3}{c}{(2 - 12)} & \multicolumn{3}{c}{(4 - 14)} & \multicolumn{3}{c}{(6 - 16)} & \multicolumn{3}{c}{8 - 18} \\
					\hline
					Id & $m$ & $c$ & $\rho_{rms}$ & $m$ & $c$ & $\rho_{rms}$  & $m$ & $c$ & $\rho_{rms}$  & $m$ & $c$  & $\rho_{rms}$  & $m$ & $c$ & $\rho_{rms}$  \\
					\hline
					$\mu_2$ & $-0.51 \pm 0.01$ & $-1.37 \pm 0.14$ & $2.08$ & $-0.53 \pm 0.01$ & $-0.99 \pm 0.27$ & $1.62$ & $-0.50 \pm 0.01$ & $-1.43 \pm 0.13$ & $0.76$ & $-0.49 \pm 0.01$ & $-1.62 \pm 0.07$ & $0.39$ & $-0.48 \pm 0.01$ & $-1.71 \pm 0.02$ & $0.13$ \\
					$\mu_3$ & $-0.61 \pm 0.05$ & $-0.55 \pm 0.80$ & $19.2$ & $-0.66 \pm 0.08$ & $2.01 \pm 2.98$ & $19.7$ & $-0.55 \pm 0.01$ & $-1.09 \pm 0.50$ & $2.80$ & $-0.53 \pm 0.01$ & $-1.59 \pm 0.06$ & $0.40$ & $-0.53 \pm 0.01$ & $-1.59 \pm 0.04$ & $0.37$ \\
					$\mu_4$ & $-1.17 \pm 0.18$ & $7.97 \pm 6.13$ & $214$ & $-1.44 \pm 0.34$ & $43.1 \pm 37.9$ & $282$ & $-1.25 \pm 0.15$ & $20.6 \pm 11.9$ & $230$ & $-1.09 \pm 0.08$ & $8.89 \pm 4.06$ & $81.6$ & $-0.99 \pm 0.04$ & $4.06 \pm 1.63$ & $75.7$ \\
					$S_3$ & $0.07 \pm 0.35$ & $0.02 \pm 0.18$ & $18.7$ & $-0.37 \pm 0.58$ & $0.31 \pm 0.35$ & $17.2$ & $0.51 \pm 0.14$ & $-0.16 \pm 0.07$ & $2.31$ & $0.75 \pm 0.04$ & $-0.27 \pm 0.02$ & $0.70$ & $0.84 \pm 0.06$ & $-0.32 \pm 0.03$ & $0.87$ \\
					$S_4$ & $-3.93 \pm 1.18$ & $9.26 \pm 3.99$ & $117$ & $-6.40 \pm 2.87$ & $22.3 \pm 14.5$ & $52.3$ & $-4.72 \pm 1.08$ & $13.6 \pm 4.76$ & $62.4$ & $-3.67 \pm 0.51$ & $9.01 \pm 2.00$ & $110$ & $-2.99 \pm 0.28$ & $6.44 \pm 1.00$ & $41.2$ \\
					\hline \hline
				\end{tabular}}
				\label{tab: mcvalstophat}
			\end{center}
		\end{table*}

\section{Aperture mass filter}

Table\,\ref{tab: mcvalsapmass} is the same as Table\,\ref{tab: mcvalstophat} but assuming an aperture mass filter is used to smooth the shear field. Contrary to the previous two cases, it turns out that Eq.(\ref{eq: obsvsth}) does not hold anymore when noise is added to the simulated maps. Indeed, $m$ can become larger and negative or the fit residuals become very large or the fitting procedure do not converge at all (which explains the missing rows in the denoising case). It is not immediate to understand such unexpected results. We nevertheless note that the aperture mass filter is typically use to smooth out all features in the convergence map in order to highlight the presence of statistically meaningful peaks. As such, we expect quite small and roughly constant values of the skewness and kurtosis which can hardly be fitted by a linear relation. This is indeed what we qualitatively find although we have not performed any attempt to quantitatevely validate this hypothesis.
  We therefore only conclude here that the aperture mass must not be used when measuring higher order convergence moments. 
		
		\begin{table*}
			\begin{center}
				\caption{Calibration parameters, $(m, c)$, and rms percentage residuals, $\rho_{rms} = rms[100 \times (1 - \mu_n(fit)/\mu_n(obs))]$, for different moments. Note that $c$ is in units of $(10^{-6}, 10^{-9}, 10^{-11})$ for $(\mu_2, \mu_3, \mu_4)$, respectively. An Aperture Mass filter is used to smooth the map.} 
				\resizebox{17cm}{!}{
					\begin{tabular}{cccccccccccccccc}
						\hline \hline
						\multicolumn{16}{c}{No mass reconstruction - No noise - No denoising} \\
						\hline
						$\theta_0$ range & \multicolumn{3}{c}{(2 - 20)} & \multicolumn{3}{c}{(2 - 12)} & \multicolumn{3}{c}{(4 - 14)} & \multicolumn{3}{c}{(6 - 16)} & \multicolumn{3}{c}{8 - 18} \\
						\hline
						Id & $m$ & $c$ & $\rho_{rms}$ & $m$ & $c$ & $\rho_{rms}$  & $m$ & $c$ & $\rho_{rms}$  & $m$ & $c$  & $\rho_{rms}$  & $m$ & $c$ & $\rho_{rms}$  \\
						\hline
						$\mu_2$ & $-0.56 \pm 0.05$ & $0.49 \pm 0.09$ & $4.46$ & 
						$-0.66 \pm 0.08$ & $0.69 \pm 0.17$ & $3.41$  & $-0.52 \pm 0.05$ & $0.47 \pm 0.08$ & $1.70$ & $-0.45 \pm 0.02$ & $0.35 \pm 0.03$ & $0.78$ & $-0.40 \pm 0.01$ & $0.29 \pm 0.02$ & $0.40$ \\
						$\mu_3$ & $2.07 \pm 0.14$ & $0.30 \pm 0.05$ & $4.26$ & $1.76 \pm 0.20$ & $0.47 \pm 0.10$ & $2.93$ & $2.10 \pm 0.16$ & $0.34 \pm 0.06$ & $2.04$ & $2.38 \pm 0.09$ & $0.26 \pm 0.02$ & $0.97$ & $2.53 \pm 0.06$  & $0.22 \pm 0.01$ & $0.60$ \\
						$\mu_4$ & $4.62 \pm 0.33$ & $0.28 \pm 0.03$ & $6.01$ & $4.00 \pm 0.43$ & $0.42 \pm 0.09$ & $3.92$ & $4.39 \pm 0.38$ & $0.35 \pm 0.06$ & $3.29$ & $5.06 \pm 0.25$ & $0.27 \pm 0.02$ & $1.90$ & $5.50 \pm 0.21$ & $0.24 \pm 0.02$ & $1.44$ \\
						$S_3$ & $8.40 \pm 1.90$ & $-0.55 \pm 0.32$ & $5.11$ & $11.2 \pm 5.92$ & $-1.06 \pm 1.04$ & $5.24$ & $8.37 \pm 1.12$ & $-0.58 \pm 0.19$ & $1.65$ & $6.76 \pm 0.55$ & $-0.32 \pm 0.09$ & $0.89$ & $5.88 \pm 0.33$ & $-0.18 \pm 0.05$ & $0.56$ \\
						$S_4$ & $10.4 \pm 3.04$ & $0.99 \pm 1.49$ & $7.72$ & $15.4 \pm 10.2$ & $-1.90 \pm 5.83$ & $9.18$ & $10.7 \pm 1.91$ & $0.51 \pm 1.03$ & $2.32$ & $8.23 \pm 0.90$ & $1.77 \pm 0.46$ & $1.19$ & $6.91 \pm 0.50$ & $2.40 \pm 0.23$ & $0.64$ \\
						\hline \hline
						\multicolumn{16}{c}{Yes mass reconstruction - No noise - No denoising} \\
						\hline
						$\theta_0$ range & \multicolumn{3}{c}{(2 - 20)} & \multicolumn{3}{c}{(2 - 12)} & \multicolumn{3}{c}{(4 - 14)} & \multicolumn{3}{c}{(6 - 16)} & \multicolumn{3}{c}{8 - 18} \\
						\hline
						Id & $m$ & $c$ & $\rho_{rms}$ & $m$ & $c$ & $\rho_{rms}$  & $m$ & $c$ & $\rho_{rms}$  & $m$ & $c$  & $\rho_{rms}$  & $m$ & $c$ & $\rho_{rms}$  \\
						\hline
						$\mu_2$ & $-0.48 \pm 0.12$ & $0.54 \pm 0.21$ & $8.86$ & $-0.71 \pm 0.25$ & $1.05 \pm 0.52$ & $8.05$ & $-0.35 \pm 0.08$ & $0.44 \pm 0.14$ & $2.57$ & $-0.21 \pm 0.04$ & $0.23\pm 0.06$ & $1.07$ & $-0.14 \pm 0.02$ & $0.14 \pm 0.03$ & $0.52$ \\
						$\mu_3$ & $3.47 \pm 0.87$ & $0.60 \pm 0.31$ & $15.6$ & $1.92 \pm 1.95$ & $1.57 \pm 1.12$ & $14.6$ & $5.15 \pm 0.52$ & $0.44 \pm 0.19$ & $3.62$ & $6.05 \pm 0.19$ & $0.19 \pm 0.05$ & $1.15$ & $6.38 \pm 0.09$ & $0.11 \pm 0.02$ & $0.49$  \\
						$\mu_4$ & $6.63 \pm 1.57$ & $0.57 \pm 0.19$ & $19.9$ & $4.00 \pm 2.94$ & $1.38 \pm 0.78$ & $17.5$ & $9.62 \pm 1.08$ & $0.61 \pm 0.16$ & $5.14$ & $11.7 \pm 0.56$ & $0.40 \pm 0.05$ & $2.17$ & $12.8 \pm 0.52$ & $0.31 \pm 0.03$ & $1.69$ \\
						$S_3$ & $12.2 \pm 0.75$ & $-0.82 \pm 0.13$ & $2.28$ & $14.4 \pm 1.98$ & $-1.21 \pm 0.35$ & $1.34$ & $13.0 \pm 0.91$ & $-0.98 \pm 0.16$ & $1.10$ & $11.4 \pm 0.59$ & $-0.72 \pm 0.10$ & $0.73$ & $10.4 \pm 0.39$ & $-0.55 \pm 0.06$ & $0.50$ \\
						$S_4$ & $15.2 \pm 1.68$ & $0.55 \pm 0.88$ & $3.36$ & $17.7 \pm 6.84$ & $-0.91 \pm 4.07$ & $3.49$ & $15.4 \pm 1.44$ & $0.15 \pm 0.81$ & $1.33$ & $13.4 \pm 0.75$ & $1.22 \pm 0.38$ & $0.77$ & $12.3 \pm 0.32$ & $1.74 \pm 0.15$ & $0.33$ \\
						\hline \hline
			
							\multicolumn{16}{c}{No mass reconstruction - Yes noise - No denoising} \\
							\hline
							$\theta_0$ range & \multicolumn{3}{c}{(2 - 20)} & \multicolumn{3}{c}{(2 - 12)} & \multicolumn{3}{c}{(4 - 14)} & \multicolumn{3}{c}{(6 - 16)} & \multicolumn{3}{c}{8 - 18} \\
							\hline
							Id & $m$ & $c$ & $\rho_{rms}$ & $m$ & $c$ & $\rho_{rms}$  & $m$ & $c$ & $\rho_{rms}$  & $m$ & $c$  & $\rho_{rms}$  & $m$ & $c$ & $\rho_{rms}$  \\
							\hline
							$\mu_2$ & $-0.29 \pm 0.18$ & $-0.33 \pm 0.32$ & $19.0$ & $-0.07 \pm 0.42$ & $-0.83 \pm 0.88$ & $19.6$ & $-0.54 \pm 0.04$ & $-0.05 \pm 0.07$ & $2.71$ & $-0.60 \pm 0.01$ & $0.05 \pm  0.02$ & $0.78$ & $-0.62 \pm 0.01$ & $0.08 \pm 0.01$ & $0.21$ \\
							$\mu_3$ & $-0.21 \pm 0.04$ & $0.07 \pm 0.01$ & $4.34$ & $-0.28 \pm 0.06$ & $0.11 \pm 0.03$ & $3.10$ & $-0.21 \pm 0.04$ & $0.08 \pm 0.01$ & $1.93$ & $-0.14 \pm 0.02$ & $0.06 \pm 0.01$ & $0.99$ & $-0.11 \pm 0.02$ & $0.05 \pm 0.01$ & $0.79$ \\
							$\mu_4$ & $11.17 \pm 7.62$ & $-0.53 \pm 0.87$ & $264$ & $17.4 \pm 17.1$ & $-2.21 \pm 3.85$ & $274$ & $1.88 \pm 0.83$ & $-0.10 \pm 0.13$ & $23.7$ & $0.77 \pm 0.16$ & $0.02 \pm 0.02$ & $5.15$ & $0.53 \pm 0.03$ & $0.03 \pm 0.01$ & $0.98$ \\
							$S_3$ & $5.28 \pm 1.20$ & $-0.36 \pm 0.20$ & $4.50$ & $6.89 \pm 3.57$ & $-0.65 \pm 0.63$ & $4.39$ & $5.39 \pm 0.66$ & $-0.40 \pm 0.11$ & $1.48$ & $4.43 \pm 0.38$ & $-0.24 \pm 0.06$ & $0.90$ & $3.80 \pm 0.23$ & $-0.14 \pm 0.04$ & $0.58$ \\
							$S_4$ & $4.43 \pm 1.31$ & $2.05 \pm 0.66$ & $5.32$ & $6.59 \pm 5.08$ & $0.77 \pm 2.98$ & $6.43$ & $4.85 \pm 1.04$ & $1.67 \pm 0.58$ & $1.71$ & $3.49 \pm 0.47$ & $2.38 \pm 0.24$ & $0.83
							$ & $2.87 \pm 0.16$ & $2.67 \pm 0.07$ & $0.29$ \\
							\hline \hline
							\multicolumn{16}{c}{Yes mass reconstruction - Yes noise - No denoising} \\
							\hline
							$\theta_0$ range & \multicolumn{3}{c}{(2 - 20)} & \multicolumn{3}{c}{(2 - 12)} & \multicolumn{3}{c}{(4 - 14)} & \multicolumn{3}{c}{(6 - 16)} & \multicolumn{3}{c}{8 - 18} \\
							\hline
							Id & $m$ & $c$ & $\rho_{rms}$ & $m$ & $c$ & $\rho_{rms}$  & $m$ & $c$ & $\rho_{rms}$  & $m$ & $c$  & $\rho_{rms}$  & $m$ & $c$ & $\rho_{rms}$  \\
							\hline
							$\mu_2$ & $4.75 \pm 1.93$ & $-4.35 \pm 2.20$ & $33.4$ & $10.3 \pm 5.68$ & $-12.8 \pm 8.57$ & $32.9$ & $7.12 \pm 2.47$ & $-7.91 \pm 3.39$ & $18.7$ & $5.02 \pm 1.22$ & $-5.00 \pm  1.56$ & $11.1$ & $3.69 \pm 0.67$ & $-3.33 \pm 0.79$ & $7.01$ \\
							$\mu_3$ & $0.51 \pm 0.35$ & $0.05 \pm 0.06$ & $167$ & $-0.17 \pm 0.96$ & $0.29 \pm 0.32$ & $163$ & $0.34 \pm 0.31$ & $0.12 \pm 0.09$ & $16.2$ & $0.64 \pm 0.17$ & $0.03 \pm 0.04$ & $5.11$ & $0.79 \pm 0.08$ & $ 0.01 \pm  0.02$  & $1.88$ \\
							$\mu_4$ & $44.9 \pm 19.9$ & $-1.02 \pm 0.89$ & $48.8$ & $111 \pm 77.8$ & $-7.97 \pm 8.46$ & $51.5$ & $78.8 \pm 38.9$ & $-4.06 \pm  3.31$ & $37.6$ & $57.2 \pm 18.8$ & $-2.07 \pm 1.27$ & $25.3$ & $42.7 \pm 9.51$ & $-1.07 \pm 0.51$ & $16.6$ \\
							$S_3$ & $-3.84 \pm 0.87$ & $0.52 \pm 0.16$ & $218$ & $-3.59 \pm 2.10$ & $0.48 \pm 0.38$ & $255$ & $-4.04 \pm 0.14$ & $0.57 \pm 0.02$ & $4.88$ & $-3.75 \pm 0.22$ & $0.52 \pm 0.04$ & $4.83$ & $-3.40 \pm 0.21$ & $0.47 \pm 0.03$ & $3.46$ \\
							$S_4$ & $-1.71 \pm 0.12$ & $3.49 \pm 0.07$ & $0.77$ & $-1.53 \pm 0.16$ & $3.38 \pm 0.10$ & $0.55$ & $-1.75 \pm 0.10$ & $3.52 \pm 0.06$ & $0.33$ & $-1.92 \pm 0.05$ & $3.61 \pm 0.03$ & $0.15$ & $-1.98 \pm 0.03$ & $3.64 \pm 0.02$ & $0.09$ \\
							\hline \hline
				
								\multicolumn{16}{c}{No mass reconstruction - Yes noise - Yes denoising} \\
								\hline
								$\theta_0$ range & \multicolumn{3}{c}{(2 - 20)} & \multicolumn{3}{c}{(2 - 12)} & \multicolumn{3}{c}{(4 - 14)} & \multicolumn{3}{c}{(6 - 16)} & \multicolumn{3}{c}{8 - 18} \\
								\hline
								Id & $m$ & $c$ & $\rho_{rms}$ & $m$ & $c$ & $\rho_{rms}$  & $m$ & $c$ & $\rho_{rms}$  & $m$ & $c$  & $\rho_{rms}$  & $m$ & $c$ & $\rho_{rms}$  \\
								\hline
								$\mu_2$ & $-0.86 \pm 0.06$ & $0.33 \pm 0.10$ & $10.2$ & $-0.95 \pm 0.12$ & $0.54 \pm 0.24$ & $9.79$ & $-0.80 \pm  0.03$ & $0.28 \pm 0.05$ & $2.47$ & $-0.74 \pm 0.01$ & $0.19 \pm 0.02$ & $0.93$ & $-0.72 \pm 0.01$ & $0.16 \pm 0.01$ & $0.58$ \\
								$\mu_3$ & $-0.21 \pm 0.04$ & $0.07 \pm 0.01$ & $4.34$ & $-0.28 \pm 0.06$ & $0.11 \pm 0.03$ & $3.10$ & $-0.21 \pm 0.04$ & $0.08 \pm 0.01$ & $1.93$ & $-0.14 \pm 0.02$ & $0.06 \pm 0.01$ & $0.99$ & $-0.11 \pm 0.02$ & $0.05 \pm 0.01$ & $0.79$ \\
								$\mu_4$ & $-23.2 \pm 15.6$ & $1.36 \pm 2.00$ & $3255$ & $-36.2 \pm 34.4$ & $4.89 \pm 7.92$ & $3907$ & $-3.69 \pm 1.75$ & $0.41 \pm 0.28$ & $252$ & $-1.23 \pm 0.43$ & $0.13 \pm 0.05$ & $100$ & $-0.54 \pm 0.18$ & $0.07 \pm 0.01$ & $9.27$ \\
															\hline \hline
								\multicolumn{16}{c}{Yes mass reconstruction - Yes noise - Yes denoising} \\
								\hline
								$\theta_0$ range & \multicolumn{3}{c}{(2 - 20)} & \multicolumn{3}{c}{(2 - 12)} & \multicolumn{3}{c}{(4 - 14)} & \multicolumn{3}{c}{(6 - 16)} & \multicolumn{3}{c}{8 - 18} \\
								\hline
								Id & $m$ & $c$ & $\rho_{rms}$ & $m$ & $c$ & $\rho_{rms}$  & $m$ & $c$ & $\rho_{rms}$  & $m$ & $c$  & $\rho_{rms}$  & $m$ & $c$ & $\rho_{rms}$  \\
								\hline
								$\mu_2$ & $-0.66 \pm 0.03$ & $0.16 \pm 0.05$ & $9.13$ & $-0.73 \pm 0.07$ & $0.28 \pm 0.13$ & $9.63$ & $-0.67 \pm 0.04$ & $0.19 \pm 0.06$ & $2.53$ & $-0.62 \pm 0.01$ & $0.12 \pm 0.02$ & $0.69$ & $-0.60 \pm 0.01$ & $0.10 \pm 0.01$ & $0.27$ \\
								$\mu_3$ & $0.51 \pm 0.35$ & $0.05 \pm 0.06$ & $167$ & $-0.17 \pm 0.96$ & $0.29 \pm 0.32$ & $163$ & $0.34 \pm 0.31$ & $0.12 \pm 0.09$ & $16.2$ & $0.64 \pm 0.17$ & $0.03 \pm 0.04$ & $5.11$ & $0.79 \pm 0.08$ & $0.01 \pm 0.02$ & $1.88$ \\
								$\mu_4$ & $-47.9 \pm 25.1$ & $1.26 \pm 0.99$ & $57.4$ & $-165 \pm 126$ & $12.5 \pm 13.1$ & $55.0$ & $-107 \pm 59.9$ & $5.92 \pm 4.69$ & $43.6$ & $-72.7 \pm 29.3$ & $2.98 \pm 1.79$ & $33.6$ & $-51.0 \pm 15.1$ & $1.57 \pm 0.74$ & $26.2$ \\
																\hline \hline
							\end{tabular}}
							\label{tab: mcvalsapmass}
						\end{center}
					\end{table*}	

\begin{thebibliography}{99}

\bibitem[\protect\citeauthoryear{Abbott et al.}{2015}]{Abb15}
Abbott, T., Abdalla, F.B., Allam, S., Amara, A., Annis, J., et al. 2015, preprint arXiv\,:1507.05552 

\bibitem[\protect\citeauthoryear{Bartelmann \& Schneider}{2001}]{B&S2001}
Bartelmann, M., Schneider, P. 2001, PhR, 340, 291

\bibitem[\protect\citeauthoryear{Becker et al. }{2015}]{DES2015}
Becker, M.R., Troxel, M.A., MacCrann, N., Krause, E., Eifler, T.F., et al. 2015, preprint arXiv\,:1507.05598

\bibitem[\protect\citeauthoryear{Bernardeau et al.}{1997}]{Berny1997}
Bernardeau, F., Van Waerbeke, L., Mellier, Y. 1997, A\&A, 322, 1

\bibitem[\protect\citeauthoryear{Blanton et al.}{2003}]{Blan2003}
Blanton, M.R., Hogg, D.W., Bahcall, N.A., Brinkmann, J., Britton, M. et al. 2003, Apj,  592, 819

\bibitem[\protect\citeauthoryear{Blanton et al.}{2005a}]{Blan2005a}
Blanton, M.R., Lupton, R.H., Schlegel, D.J., Strauss, M.A., Bronkmann J., Fukugita M., Loveday, J. 2005a, ApJ, 631, 208

\bibitem[\protect\citeauthoryear{Blanton et al.}{2005b}]{Blan2005b}
Blanton, M.R., Schlegel, D.J., Strauss, M.A., Brinkmann, J., Finkbeiner, D. et al., 2005b, ApJ, 129, 2562

\bibitem[\protect\citeauthoryear{Carretero et al.}{2015}]{Carr2015}
Carretero, J., Castander F.J., Gazta\~{n}aga, E., Crocce, M., Fosalba, P., 2015, MNRAS 447, 646

\bibitem[\protect\citeauthoryear{Crocce et al.}{2015}]{Crocce2015b}
Crocce, M., Castander F.J., Gazta\~{n}aga, E., Fosalba, P., Carretero, J. 2015, MNRAS, 453, 1513

\bibitem[\protect\citeauthoryear{de Jong et al. }{2015}]{Kids}
de Jong, J.T.A., Verdoes Kleijn, G.A., Boxhoorn, D.R., Buddelmeijer, H., Capaccioli, M., et al. 2015, A\&A, 582, 62

\bibitem[\protect\citeauthoryear{Fosalba et al.}{2015a}]{Fos2015a}
Fosalba, P., Crocce, M., Gazta\~{n}aga, E., Castander, F.J. 2015a, MNRAS, 448, 2987

\bibitem[\protect\citeauthoryear{Fosalba et al.}{2015b}]{Fos2015b}
Fosalba, P., Gazta\~{n}aga, E., Castander, F.J., Crocce,M. 2015b, MNRAS, 447, 1319

\bibitem[\protect\citeauthoryear{Green et al.}{2012}]{WFIRST2012}
Green et al. 2012, arXiv:1208.4012

\bibitem[\protect\citeauthoryear{Giocoli et al.}{2015}]{Gio15}
Giocoli, C., Jullo, E., Metcalf, B., de la Torre, S., Yepes, G., Prada, F., et al. 2015, preprint arXiv\,:1511.08211

\bibitem[\protect\citeauthoryear{Hamana et al.}{2004}]{Hama2004}
Hamana, T., Takada, M., Yoshida, N., 2004, MNRAS, 350, 893

\bibitem[\protect\citeauthoryear{Heymans et al.}{2012}]{H12}
Heymans, C., van Waerbeke, L., Miller, L., Erben, T., Hildebrandt, H., et al. 2012, MNRAS, 427, 146

\bibitem[\protect\citeauthoryear{Hildebrandt et al.}{2016}]{H16}
Hildebrandt, H., Choi, A., Heymans, C., Blake, C., Erben, T., et al. 2016, preprint arXiv\,:1603.07722

\bibitem[\protect\citeauthoryear{Hoffmann et al.}{2015}]{Hoff15}
Hoffmann, K., Bel, J., Gazta\~{n}agaga, E., Crocce, M., Fosalba, P., Castander, F.J. 2015, MNRAS, 447, 1724

\bibitem[\protect\citeauthoryear{Jain \& Seljak}{1997}]{JaSel1997}
Jain, B., Seljak, U. 1997, ApJ, 484, 560

\bibitem[\protect\citeauthoryear{Jain, Seliak \& White}{2000}]{JSW2000}
Jain, B., Seljak, U., White, S. 2000, ApJ, 530, 547

\bibitem[\protect\citeauthoryear{Jee et al.}{2013}]{Jee13}
Jee, M.J., Tyson, J.A., Schneider, M.D., Wittman, D., Schmidt, S., Hilbert, S. 2013, ApJ, 765, 74

\bibitem[\protect\citeauthoryear{Jee et al.}{2015}]{Jee15}
Jee, M.J., Tyson, J.A., Hilbert, S., Schneider, M.D., Schmidt, S., Wittman, D. 2015, preprint arXiv\,:1510.03962

\bibitem[\protect\citeauthoryear{Kaiser \& Squires}{1993}]{KS93}
Kaiser, N., Squires, G. 1993, ApJ, 404, 441 

\bibitem[\protect\citeauthoryear{Kilbinger et al.}{2013}]{Kil13}
Kilbinger, M., Fu, L., Heymans, C., Simpson, F., Benjamin, J., et al. 2013, MNRAS, 430, 2200

\bibitem[\protect\citeauthoryear{Kilbinger}{2014}]{Kilby2014}
Kilbinger, M. 2014, arXiv:1411.0115

\bibitem[\protect\citeauthoryear{Kitching et al. }{2014}]{Tom14}
Kitching, T.D., Heavens, A.F., Alsing, J., Erben, T., Heymans, C., et al. 2014, MNRAS, 442, 1326 

\bibitem[\protect\citeauthoryear{Laureijs et al.}{2011}]{Laur2011}
Laureijs et al. 2011, arXiv:1110.3193

\bibitem[\protect\citeauthoryear{LSST}{2009}]{LSST2009}
LSST Science Collaboration et al. 2009, arXiv:0912.0201

\bibitem[\protect\citeauthoryear{Munshi}{2000}]{M2000}
Munshi, D. 2000, MNRAS, 318, 145

\bibitem[\protect\citeauthoryear{Munshi \& Jain}{2000}]{MJ2000}
Munshi, D., Jain, B. 2000, MNRAS, 318, 109

\bibitem[\protect\citeauthoryear{Munshi et al.}{2001}]{MJ01}
Munshi, D., Jain, B. 2001, MNRAS, 322, 107

\bibitem[\protect\citeauthoryear{Munshi \& Jain}{2001}]{MJ2001}
Munshi, D., Jain, B. 2001, MNRAS, 322, 107

\bibitem[\protect\citeauthoryear{Munshi et al.}{2008}]{Mun2008}
Munshi, D., Valageas, P., Van Waerbeke, L., Heavens, A. 2008, PhR, 462, 67

\bibitem[\protect\citeauthoryear{Pratten et al.}{2016}]{Pr16}
Pratten, G., Munshi, D., Valageas, P., Brax, P. 2016, preprint arXiv\,:1602.06711

\bibitem[\protect\citeauthoryear{Takada \& White}{2004}]{TW2004}
Takada, M., White, M. 2004, ApJ, 601, L1

\bibitem[\protect\citeauthoryear{Takada \& Jain}{2004}]{TJ2004}
Takada, M., Jain, B. 2004, MNRAS, 347, 897

\bibitem[\protect\citeauthoryear{Valageas}{2000}]{V2000}
Valageas, P. 2000, A \& A, 356, 771

\bibitem[\protect\citeauthoryear{Valageas, Munshi \& Barber}{2005}]{VMB2005}
Valageas, P., Munshi, D., Barber, A.J. 2005, MNRAS, 356, 386

\bibitem[\protect\citeauthoryear{Van Waerbeke et al.}{2013}]{Ludo2013}
van Waerbeke, L., Benjamin, J., Erben, T., Heymans, C., Hildebrandt, H. et al. 2013, MNRAS, 433, 3373

\bibitem[\protect\citeauthoryear{Zehavi et al.}{2011}]{Zeha2011}
Zehavi, I., Zheng, Z., Weinberg, D.H., Blanton, M.R., Bahcall, N.A.  et al. 2011, ApJ, 736, 59

\end{thebibliography}
\end{document}